%% file: MAIN_submission.tex
\documentclass[manuscript]{acmart}

\AtBeginDocument{%
  }

\setcopyright{acmlicensed}
\copyrightyear{2018}
\acmYear{2018}
\acmDOI{XXXXXXX.XXXXXXX}

\acmConference[Conference acronym 'XX]{Make sure to enter the correct
  conference title from your rights confirmation emai}{June 03--05,
  2018}{Woodstock, NY}
\acmISBN{978-1-4503-XXXX-X/18/06}

\usepackage{subfiles}
\usepackage{subfigure}

\usepackage{array}
\usepackage{pdflscape}
\usepackage{longtable}
\usepackage{multicol}
\usepackage{caption}
\usepackage{multirow}
\usepackage[inline]{enumitem}

\newcolumntype{L}[1]{>{\raggedright\let\newline\\\arraybackslash\hspace{0pt}}m{#1}}
\newcolumntype{C}[1]{>{\centering\let\newline\\\arraybackslash\hspace{0pt}}m{#1}}
\newcolumntype{R}[1]{>{\raggedleft\let\newline\\\arraybackslash\hspace{0pt}}m{#1}}

\newlist{customlist}{enumerate}{1}
\setlist[customlist,1]{
  label=\arabic*., 
  labelsep=0.5em, 
  left=0pt, 
  itemsep=0pt, 
  parsep=0pt, 
  before=, 
  after=
}

\begin{document}

\title{Socially Constructed Treatment Plans: Analyzing Online Peer Interactions to Understand How Patients Navigate Complex Medical Conditions}

\author{Madhusudan Basak}
\email{madhusudan.basak.gr@dartmouth.edu}
\affiliation{%
  \institution{Dartmouth College}
  \city{Hanover}
  \state{New Hampshire}
  \country{United States}
}
\affiliation{%
  \institution{BUET}
  \city{Dhaka}
  \state{Dhaka}
  \country{Bangladesh}
}
\author{Omar Sharif}
\affiliation{%
  \institution{Dartmouth College}
  \city{Hanover}
  \state{New Hampshire}
  \country{United States}
}

\author{Jessica Hulsey}
\email{jhulsey@addictionpolicy.org}
\affiliation{%
  \institution{Addiction Policy Forum}
  \city{Bethesda}
  \state{Maryland}
  \country{United States}
}

\author{Elizabeth C. Saunders}
\email{elizabeth.c.saunders@dartmouth.edu}
\affiliation{%
  \institution{Dartmouth College}
  \city{Hanover}
  \state{New Hampshire}
  \country{United States}
}

\author{Daisy J. Goodman}
\email{daisy.j.goodman@dartmouth.edu}
\affiliation{%
  \institution{Dartmouth College}
  \city{Hanover}
  \state{New Hampshire}
  \country{United States}
}

\author{Luke J. Archibald}
\email{Luke.J.Archibald@Dartmouth.edu}
\affiliation{%
\institution{Dartmouth College}
  \city{Hanover}
  \state{New Hampshire}
  \country{United States}
}

\author{Sarah M. Preum}
\email{Sarah.masud.preum@dartmouth.edu}
\affiliation{%
  \institution{Dartmouth College}
  \city{Hanover}
  \state{New Hampshire}
  \country{United States}
}

\renewcommand{\shortauthors}{Basak et al.}

\begin{abstract}
  When faced with complex and uncertain medical conditions (e.g., cancer, mental health conditions, recovery from substance dependency), millions of patients seek online peer support. In this study, we leverage content analysis of online discourse and ethnographic studies with clinicians and patient representatives to characterize how treatment plans for complex conditions are “socially constructed.” Specifically, we ground online conversation on medication-assisted recovery treatment to medication guidelines and subsequently surface when and why people deviate from the clinical guidelines. We characterize the implications and effectiveness of socially constructed treatment plans through in-depth interviews with clinical experts. Finally, given the enthusiasm around AI-powered solutions for patient communication, we investigate whether and how socially constructed treatment-related knowledge is reflected in a state-of-the-art large language model (LLM). Leveraging a novel mixed-method approach, this study highlights critical research directions for patient-centered communication in online health communities. 
  


\end{abstract}


\keywords{Computer Mediated Communication, Health - Clinical, Social Media/Online Communities, HCI for Development, Empirical study that tells us about people, Content Analysis, Conversation Analysis}


\renewcommand{\authorsaddresses}{}
\maketitle

\subfile{Sections/1_Introduction}
\subfile{Sections/2_Related_Works}
\subfile{Sections/3_Methods}

\subfile{Sections/4_Treatment_Plan_Aspects}
\subfile{Sections/5_Peer_Suggested_Strategies}

\subfile{Sections/6_LLMs_Responses}
\subfile{Sections/7_Interviews}
\subfile{Sections/8_Discussion}
\subfile{Sections/9_Limitations_and_Future_Works}
\subfile{Sections/10_Conclusion}

\bibliographystyle{ACM-Reference-Format}
\bibliography{Ref}



\end{document}

%% file: Sections/1_Introduction.tex
\section{Introduction}
A recent study found that approximately 70\% of the US population resorts to online communities~\cite{kanchan2023social} to connect with their peers and exchange information to manage various aspects of their lives including education, entertainment, health, and profession. One of the most important fields in which online communities have a significant impact is healthcare~\cite{househ2013use}. People navigate online health communities to deal with uncertain and complex medical conditions like mental health conditions, chronic and acute conditions like cancer, and recovery from substance use disorder. These communitiess serve as a major resource for assessing their conditions and obtaining relevant information \cite{10.1145/2702123.2702566}. These complex treatment conditions present challenges where treatment knowledge is not universal but highly situational, influenced by individual circumstances, experiences, and responses to treatment. In these online spaces, patients engage in the social construction of treatment plans, which may vary significantly from formal clinical guidelines, shaped by the diversity of their lived experiences and shared insights.

We explore the dynamic nature of treatment plans in online health communities, focusing on how peers socially construct them and how they differ from clinical guidelines. We also explore how knowledge is situated and contested within these communities, examining how patients’ perspectives evolve through peer exchanges. A better knowledge-creation process can aid patients in making informed decisions, while the prevalence of unverified rumors or misinformation can be consequential and harmful. Studying the knowledge-generation process related to treatment plans in online communities can provide insights to address concerns that impact treatment adherence and violate health safety.

We chose Reddit as the platform for this study due to its rich source of user-generated content where patients engage in meaningful discussions. Reddit's community-driven structure allows for the natural development of treatment narratives, offering a valuable lens through which we can explore the social construction of treatment plans in online health communities. By examining Medication for Opioid Use Disorder (MOUD) as a use case, we aim to understand the ways in which individuals dealing with opioid addiction navigate treatment decisions, use peer support advice and shared knowledge in online spaces.

In this work, we aim to demonstrate how patients construct categories, norms, and meanings related to their treatment plans in online health communities. We seek to understand how treatment plans are socially constructed and how they deviate from or align with formal clinical guidelines. An important aspect of our study is to evaluate how clinicians view the safety and effectiveness of these socially constructed treatment plans. These objectives guide us toward answering the following research questions.

\begin{quote}
    \textbf{RQ1: What are the aspects of the treatment plans socially constructed on online health communities?}\\
    \textbf{RQ2: How do peers address these issues? What are the characteristics of the knowledge constructed through the peer interaction?}
\end{quote}

We also note that large language models (LLMs), particularly interactive conversational chatbot versions like ChatGPT \cite{openai2024gpt4technicalreport}, have recently become popular for obtaining information online alongside online communities. Given the widespread popularity of LLMs in content analysis and integration with online platforms, including search engines and social media \cite{openai_reddit_partnership_2024, vaishya2023chatgpt}, we speculate that they could be used in online health communities. While LLMs excel at capturing popular suggestions, there remains a risk of missing rare or less frequent concepts. Their role in reflecting complex, socially constructed treatment knowledge requires further exploration. This observation leads us to the third research question, where we explored the scope and limitations of LLMs in reflecting socially constructed knowledge by comparing their strategies with those offered by peers’ comments in response to posts.


\begin{quote}
    \textbf{RQ3: How well can large language models reflect socially constructed knowledge related to treatment plans?}
\end{quote}

To answer these three questions, we performed a mixed-method study of posts, comments, and LLM responses and conducted semi-structured interviews with experts to validate our methods and results at various stages of the study. We found that the socially constructed treatment plans for individuals with MOUD are highly complex, with evidence of knowledge gaps and false information that can lead to rumors and misinformation. These findings raise new questions and areas for clinical research. Finally, while LLMs offer a more reliable information space, they cannot capture all aspects of the socially constructed knowledge related to treatment. Although our study focuses on MOUD treatment, our methods are generalizable to other medical conditions and underprivileged communities.






%% file: Sections/2_Related_Works.tex
\section{Related Works}

\subsection{Social Constructionism in Healthcare and HCI Scholarship}

Social constructionism is a theoretical approach that examines how knowledge and understanding of the world are created through social interactions and shared meanings \cite{andrews2012social}. Understanding is historically and culturally specific, meaning concepts and categories can change over time and across different societies \cite{maines2000social, pinch1984social}. Knowledge and social action are interlinked, with our understanding of the world influencing how we behave and interact with others. When applied to healthcare, it offers insights into how medical knowledge, illness experiences, and healthcare practices are shaped by social and cultural factors. Social constructionism has been applied to various aspects of healthcare, including understanding chronic illness experience \cite{martin2009social, charmaz1990discovering}. Martin et al. discuss the creative tensions between empirical and intellectual critique in managing chronic illnesses in primary care \cite{martin2009social}. Brown et al. examine how medical diagnoses and illnesses are socially constructed \cite{brown1995naming}. Another study explores how cultural and social factors shape medical knowledge and practices \cite{lupton2012medicine}. Conrad et al. discuss how illnesses are socially constructed and the implications for healthcare policy \cite{conrad2010social}. A recent study applies social constructionism to understand dementia care experiences \cite{farhana2023social}. In this work, we extend the concept of social construction for treatment plan, i.e., how peers socially construct treatment plan to navigate uncertain and complex medical conditions.



\subsection{Collective Sensemaking in Online Communities}

Online communities have become critical spaces for information dissemination and knowledge generation. Knowledge is generated in these communities through participation, where users engage in sharing personal experiences, asking questions, and responding to others. This iterative process allows a constant refinement of interpretations and solutions, contributing to a dynamic and evolving body of collective knowledge \cite{10.1145/3479564}. This collaborative process in which a group of individuals work together to interpret complex, ambiguous, or evolving situations is referred to as collective sensemaking. It involves the pooling of shared knowledge, experiences, and communication to create a cohesive understanding of an issue \cite{10.1145/3290605.3300563, 10.1145/3479564}. In online health communities, collective sensemaking allows users to interpret unfamiliar symptoms, navigate complex treatment options, or respond to emerging health concerns \cite{10.1145/2702123.2702566}. This process is vital for individuals dealing with health issues such as opioid use disorder (OUD), where uncertainties, stigma, and conflicting information can challenge their ability to make informed decisions about treatment \cite{alternative_chi}.

\subsection{Technology Mediated Tools for Treatment Adherence}

Treatment adherence has been a key focus in public health and healthcare innovation. Most interventions aimed at improving adherence have been developed as individual-focused, mobile health solutions. These include tools like reminder apps, digital health trackers, and telemedicine services, all designed to help patients manage their treatment regimens more effectively \cite{mistry2015technology}. Studies have examined the efficacy of these technologies in supporting adherence. Additionally, public health and human-computer interaction (HCI) research has explored user experiences with these tools \cite{pratiwi2023systematic}. However, there is a lack of research focusing on the large-scale analysis of online discourse to understand how knowledge about treatment adherence is socially constructed. Mining such discourse from online platforms could provide critical insights into sensemaking around treatment plans. 


\subsection{Medications for Opioid Use Recovery (MOUD)}

Social media has emerged as a crucial platform for sharing information, connecting with peers, and navigating various aspects of healthcare \cite{sharif2024characterizing}. People from diverse socio-demographic backgrounds increasingly turn to online communities to seek and exchange health-related information throughout different stages of their healthcare journeys. Analyzing self-reported treatment information from these communities offers a unique opportunity to uncover knowledge gaps and identify conflicting information with clinical guidelines, treatment perceptions, and misconceptions \cite{CHEN2022100061, li-etal-2014-major}. Prior studies have utilized social media data to understand various types of substance usage, including alcohol, opioids, alcohol, and others \cite{Lavertu2021.04.01.21254815}. Balsamo et al. investigated how the communities in online space can support the individual undergoing opioid usage \cite{Balsamo_Bajardi_De}. Another related study by Chancellor et al. attempted to find alternative treatment strategies by analyzing opioid-related discussions on Reddit \cite{alternative_chi}. To better understand how individuals navigate complex health conditions, in this work, we investigate online discussions regarding medications for opioid use disorder (MOUD). Opioid use disorder (OUD) is a major public health concern and leading cause of mortality in the U.S.\cite{FLORENCE2021108350}. However, widespread misconceptions, stigma, and knowledge gaps surrounding OUD treatment continue to hinder both treatment initiation and adherence.  

\subsection{LLMs for Social Knowledge Construction}

Large language models (LLMs) are ubiquitous and have shown significant performance in various syntactic, discourse, and reasoning tasks \cite{ziems-etal-2024-large}. LLMs have been particularly effective in capturing complex patterns of human interaction and making them useful for analyzing discussions around healthcare, opioid use disorder, and mental health \cite{10.1145/3613904.3642937}. Studies have shown that LLMs can identify implicit meaning, answer questions, and summarize discourse of long conversational threads \cite{ziems-etal-2024-large}. LLMs are now being used to analyze large-scale social interactions, enabling researchers to study phenomena such as collective decision-making, peer support, and the dissemination of health information. For instance, LLMs have been leveraged to identify misinformation and misconceptions shared in online health communities \cite{chen2023combatingmisinformationagellms} and have also been used to surface latent social dynamics that influence knowledge creation and dissemination \cite{choi-etal-2023-llms}. In this work, we investigate their potential to create social knowledge surrounding treatment plans regarding medications for opioid use disorder.



%% file: Sections/3_Methods.tex
\section{Methods}

In this work, we considered online health community discourse on MOUD to explore how the MOUD treatment plan for recovery is socially constructed. We analyzed the posts and comments from the Reddit Sub-community ``r/Suboxone" as a use case. We considered Reddit due to the pseudo-anonymity offered by this platform, which provides the members with the unique opportunity to share their concerns and experiences without revealing their identities. This is crucial for the stigmatized communities like individuals on MOUD. We started with 1000 discourses and then performed multiple qualitative and quantitative analyses to find the answers to our research questions. The study received approval from the Institutional Review Board (IRB) at the authors’ institution.

\subsection{Background}
We are defining the terminologies specific to MOUD here. Opioid Use Disorder (OUD) refers to the distress or impairment caused by long-term dependency on opioids~\cite{dydyk2023opioid}.
Medication for opioid use
disorder (MOUD) and psychological therapy are the main two options for treating OUD~\cite{amato2011psychosocial}.
Buprenorphine and methadone are the two most popular MOUD options. Suboxone, Subutex, and Sublocade are different forms of Buprenorphine. However, these forms of Buprenorphine are medications that patients can take at home after filling in their prescriptions. Thus, there is more scope for lack of adherence as Methadone is administered at the clinic. Although this subreddit is named ``Suboxone", our subset of data also covers patients' comparative discussion on other buprenorphine options (e.g., Subutex, Sublocade), making it a rich source of data to study medication adherence and socially constructed treatment plans online communities.

\subsection{Data Collection}
We utilize the dataset curated by Sharif et al. as this is the largest online discourse dataset on the MOUD treatment annotated by main themes (they called those ``events") of recovery treatment ~\cite{sharif2024characterizing}. This dataset contains 5,000 posts labeled with five events. This dataset was collected from ``r/Suboxone," one of the most popular Reddit sub-communities, containing almost 40K members to date. The posts span from January 2, 2018, to August 6, 2022. For our analysis, we focused on 1,951 posts labeled with the event ``Taking MOUD," as it is closely tied to treatment planning. These posts include questions and concerns about medication administration, such as timing, process, dosage, frequency, and other factors affecting medication adherence. Furthermore, this data set was annotated as a multi-class, multi-label data set, facilitating the analysis of cross-sectional topics, e.g., how the medication's perceived psychological or physical effect affects adherence. We considered this event due to its high occurrence and significant direct consequences. 

\subsection{Methodological Overview}
We performed several qualitative and quantitative analyses in our study. We are briefly describing the steps here to give the reader an overview of the methodologies. We delineate the details of the analyses in the next sections. 


For RQ1, we aimed to identify the socially constructed treatment aspects from online social discourse. We randomly selected 1,000 posts from our dataset tagged with `Taking MOUD,' and performed an abductive analysis on the dataset to identify relevant themes and subthemes that represent the treatment aspects. Next, we strategically selected a subset of 100 posts, compiled a set of guideline instructions from authentic sources, and manually checked whether the contents of those posts adhere to the guidelines.

In RQ2, we performed a reflexive thematic analysis on 595 comments associated with the previously selected 100 posts and identified the strategies that represent the knowledge constructed via interactions in the community. We also performed another analysis, where we manually analyzed the comments to determine their adherence to the clinical guidelines.

As part of RQ3, we wanted to see how much socially constructed knowledge can be reflected by LLMs. As the representative of LLMs, we explored GPT-4 due to its availability and popularity. We performed mixed-method analyses here. We generated the responses of the LLM (i.e., GPT-4) using an automated way. Then we annotated each response with applicable strategies mirroring the types of strategies in RQ2. Finally, we evaluated these responses to determine whether they followed the guidelines.

We maintained continuous communication with five MOUD experts, four of whom are the authors of this paper. We verified the methods and results at step by them. We also conducted semi-structured interviews with three clinical experts on our overall findings.

\subsection{Positionality of the Authors}

The first two authors and the last author work in the intersection of natural language processing (NLP), human-computer interaction (HCI), and social computing. All three of them have extensive experience and training in working with clinical researchers. The third author is the executive director of a national nonprofit organization dedicated to thousands of US patients and their families impacted by substance use disorder. She is also involved in conducting patient-centric research to capture the lived experiences of MOUD patients. The fourth author, a Senior Clinical Research Scientist in the medical school at the authors' institution, specializes in community-engaged approaches to recovery treatment. The fifth author is a specialist in Obstetrics \& Gynecology and treating women and pregnant women with OUD. She is also the director of Women's Health Services in a Perinatal Addiction Treatment Program at a regional hospital. The sixth author is a specialist in Psychiatry and the director of the addiction treatment program at a regional hospital. The fifth and sixth authors have over 15 and 20 years of experience in treating patients with OUD and MOUD research.

%% file: Sections/4_Treatment_Plan_Aspects.tex
\section{Aspects of socially constructed treatment plan emerging from the patient narratives}
\label{aspects}

\subsection{Methods}
\label{aspects_methods}

\subsubsection{Determining treatment aspects} 
\label{subsubsec:method_aspects}
We identified the treatment aspects by abductive analysis~\cite{timmermans2012theory} of clinical guidelines of Suboxone as well as the Reddit posts seeking treatment information on Suboxone. We considered both sources to capture the hermeneutical and heuristical views in constructing the treatment plan. 

At first, we reviewed the clinical guidelines associated with buprenorphine medications (Suboxone, Subutex, and Sublocade) with guidance from our co-authors who are also experts in recovery treatment (i.e., MOUD clinicians, researchers, and patient advocates). This resulted in six high-level themes. We compiled a list of authentic sources on recovery treatment based on experts' suggestions. Some of the primary sources of clinical guidelines are listed below. 
\begin{itemize}
    \item Substance Abuse and Mental Health Services Administration (SAMHSA)~\cite{SAMSHA}: A U.S. government agency under the Department of Health and Human Services (HHS). Its main responsibility is to improve the quality and availability of substance use and mental health services in the United States.
    \item American Society of Addiction Medicine (ASAM)~\cite{ASAM}: A professional society dedicated to advancing the treatment of addiction through education, research, and advocacy. It provides a range of resources, including clinical practice guidelines.
    \item U.S. Food and Drug Administration (FDA)~\cite{FDA}: A federal agency responsible for protecting and promoting public health by regulating food safety, pharmaceuticals, medical devices, and other health-related products.
    \item Providers Clinical Support System (PCSS)~\cite{PCSS}: A national initiative designed to improve the quality of care for patients with substance use disorders.
    \item Subuxone.com~\cite{Suboxone}: An authentic website dedicated to providing information and resources about Suboxone, a prescription medication used to treat opioid use disorder. It includes information for both patients and healthcare providers.
\end{itemize}

From this list of sources, we then collected clinical instructions and grouped those according to the themes. Then, we randomly selected 1,000 posts from the 1,951 posts labeled with `Taking MOUD'. The two authors analyzed 300 posts and extracted three more themes as part of the abductive analysis. The authors discussed their findings with the experts. The experts reviewed all nine themes and refined the definitions and boundaries of these themes. In the discussion, the authors shared their experience of analyzing 300 posts and mentioned that the complexity of data could be hard to capture only by the themes. The experts agreed with the points and suggested to list out the subthemes under each theme to capture granular-level information.  

The first author then analyzed all 1,000 posts and listed down the subthemes that appeared with each theme. For example, some of the subthemes listed under the theme ``Method of administration" were ``how to dissolve" (the medication), ``problems with dissolution", ``how to swallow," ``how to spit," etc. 
The second author then reviewed the posts, and annotated themes and subthemes for additional scrutiny. Any conflicts were resolved by discussing with the experts.  

\subfile{../Tables/Stat}


\subsubsection{Align patient-narrated discourse to the guidelines:} 
\label{post-adherence-method}
Our next target was to explore whether the discussions in the posts align with the official treatment guidelines or deviate from it. For example, the proper way to use the Suboxone film is to let it dissolve, as instructed by guidelines ``While SUBOXONE sublingual film is dissolving, do not chew or swallow the film because the medicine will not work as well." Some patients mentioned they are letting it dissolve, which aligns with the guideline instruction. However, some others report that they are swallowing the film, which deviates from the guidelines.

We first strategically selected 100 posts from the previously annotated 1,000 posts to achieve this goal. The strategic selection was to ensure representative posts from each theme. Subsequently, we compare each post against the relevant clinical guidelines to determine whether the content in the post follows the clinically verified instructions. Since our list of guidelines is not exhaustive, we could not verify some of the contents. We labeled those as ``No guideline." We also considered another label ``Unable to determine," to refer to the post contents that are confusing or not relevant.

One author compared posts with guideline instructions. For each post, the author assigned one of four labels: ``Aligns with guidelines," ``Deviates from guidelines," ``No guideline," and ``Unable to determine." Particular emphasis was placed on the "deviates from guideline" label. If any part of a post contradicted the instructions, even if other parts followed them, it was labeled as "Deviates from guidelines." This approach was based on the idea that a single piece of false information can overshadow accurate details and may have more severe consequences. Another author reviewed and finalized the entire labeling process.


\subsection{Results}
\label{treatment-plan-results}
In this section, we first present the aspects of the treatment plans identified by our analysis. We then describe the contested and complementary knowledge that we found in our study. It is worth mentioning that the posts and comments used in this paper are paraphrased and sometimes shortened to protect anonymity. 

\subsubsection{Treatment aspects:} We found nine themes and 40 subthemes from our qualitative analysis of the patient narratives. These themes and subthemes represent the different aspects of the treatment plan. We are describing these treatment aspects below. The full list of themes and subthemes is available in Table~\ref{theme_subtheme_issues}. 

\textbf{Method of administration}: This theme represents the discussions on the details of the procedure of administering the medication. Different types of queries are discussed here, including how to take medications, issues with the dissolution of the medications, questions regarding swallowing and spitting, etc. We found nine subthemes under this theme. Following is an example where the post author, an individual in recovery, asks several questions, including whether to swallow the medication.


\begin{quote}

   \textbf{EP1:} \textit{  I've noticed many comments about the recommendation to spit while a film or tablet is dissolving and not to swallow it until it’s completely dissolved. I’ve used Suboxone as well as five other generic brands, and none of them included instructions about spitting until the medication is fully dissolved. I’m curious to hear your opinions on this.}
\end{quote}
Several members of the community also report issues with dissolution, as shown in the following excerpt.
\begin{quote}
    \textbf{EP2:} \textit{My Subs [Suboxone] aren't dissolving under my tongue. It should take about 10 minutes, but mine hasn't dissolved after 25 minutes. Any advice?}
\end{quote}

\textbf{Comparing different medication options:} This theme encompasses the discussions comparing different medication options such as films vs tablets, name brand vs generic brand, and medication types (e.g., Suboxone vs Methadone).  The common discussion topics are whether one option is better than another, how to switch from one to another, and the problems associated with a particular option. We found four subthemes discussed under this theme. Following is an example where the post author is comparing two forms of medications (films and tablets).

\begin{quote}
    \textbf{EP3:} \textit{I've been taking 16 mg of Suboxone (film) daily for the past 2.5 months. How can I properly request my doctor a switch to tablets instead of the films?}
\end{quote}

\textbf{Psychophysical effects
associated with medication:} People often discuss different psychophysical (physical and/or psychological) effects that happen during the recovery due to different dosing strategies. Our analysis extracted four subthemes related to psychophysical effects during recovery. In the following post, the post author wanted to know whether headache, a physical side effect, is due to taking the medication Suboxone.
\begin{quote}
  \textbf{EP4:} \textit{For the last two weeks, I’ve been suffering from severe headaches that have caused me to miss work, wake up in tears, and even try to knock myself out just to fall back asleep. I’m uncertain whether these headaches are cluster headaches or a side effect of Suboxone. Is it because I'm swallowing the Naloxone during the sleep?}
\end{quote}


\textbf{Tapering:} This theme encapsulates discussions about dosage-related questions while reducing or stopping medication intake. Five subthemes were listed for this theme. In the following post, the post author seeks information on using a substance (Xanax) during tapering.

\begin{quote}
     \textbf{EP5:} \textit{I’m working on discontinuing my use of Suboxone. Did Xanax help you mentally and temporarily during your withdrawal from Suboxone? I was taking about 0.25 mg of Xanax before bed each day—nothing excessive. Did anything else help you, or do you have any advice?}
\end{quote}


\textbf{Co-occurring alcohol
consumption:} Concurrent use of medications and alcohol is another theme that we discovered along with four subthemes. The discussions related to this theme revolves around the reason, frequency, and experience with using alcohol with medication, as depicted in the following example post.

\begin{quote}
    \textbf{EP6:} \textit{Is it possible to consume alcohol while on Suboxone? I’m aware that there are subreddits focused on alcohol consumption, but I’m curious if anyone here has used them alongside Suboxone.}
\end{quote}

\textbf{Co-occurring substance usage:} 
Although not recommended, people often use substances along with recovery medications. Some post authors want to know about reason and the consequences of concurrent substance use, while others share their practical experiences with it. We identified four subthemes in the category. Following is an example post that falls under this theme.

\begin{quote}
    \textbf{EP7:} \textit{Is it accurate that taking oxycodone or Percocet together with Suboxone, either simultaneously, on the same day, or too close together, can make you feel unwell?}
\end{quote}

\textbf{Medication during pregnancy:} We found a number of discussions where post authors share their experiences, express their concern, and  ask questions on medication during pregnancy. Five subthemes were prevalent here. Following is a post author sharing her concern about stopping the medication during pregnancy.
\begin{quote}
  \textbf{EP8:} \textit{I’ve been using Suboxone for nearly two years. My husband and I are planning to try for another baby soon, so I’ve begun tapering off. Should I completely stop before attempting to conceive again?}
\end{quote}

\textbf{Medication before surgery:} We also found people concerned about medication changes during their upcoming surgeries and asking questions relevant to it. We found two main subthemes here. Following is a post where the post author shares their concern about changes in medication for surgery.
\begin{quote}
    \textbf{EP9:} \textit{My doctor increased the dose for back pain and now plans to switch me to buprenorphine (without naloxone) a week before my surgery, so I can use Percocet afterward. I doubt two days off Suboxone is enough and question if buprenorphine will block opiates. Any advice?}
\end{quote}

\textbf{Resuming treatment after returning to the usage:} People during opioid recovery treatment often intentionally or by force of the situation go back to using opioids temporarily. It is advised to consult with the provider about how to resume the recovery treatment after the use. However, due to the unavailability of access to providers or the fear that the provider would discontinue the treatment, they seek information from their peers instead of the providers, such as when and how to resume recovery in that condition. We found four distinct subthemes under this theme. A relevant scenario is depicted in the following example.
\begin{quote}
 \textbf{EP10:} \textit{After being clean for nine months, I relapsed and used fentanyl for nine days, with seven of those days being heavy use... How long should I wait before taking Suboxone? A friend mentioned he’s okay after 24 hours, but I’ve read on Reddit that some people experience precipitated withdrawal after 80 hours.}
\end{quote}

\subfile{../Tables/Theme_subtheme_issues}

\subsubsection{Adherence to the guideline:}
\begin{figure}[h!]
  \centering
  \subfigure[Distribution of guideline alignment for posts \label{guideline-posts}]
  {\includegraphics[width=0.49\linewidth]{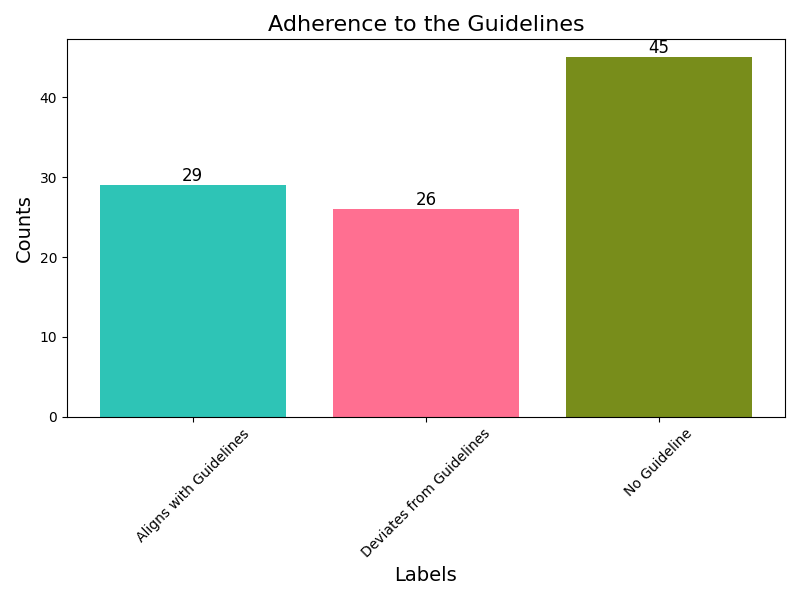}}
  \subfigure[Distribution of guideline alignment for comments \label{guideline-comments}]{\includegraphics[width=0.49\linewidth]{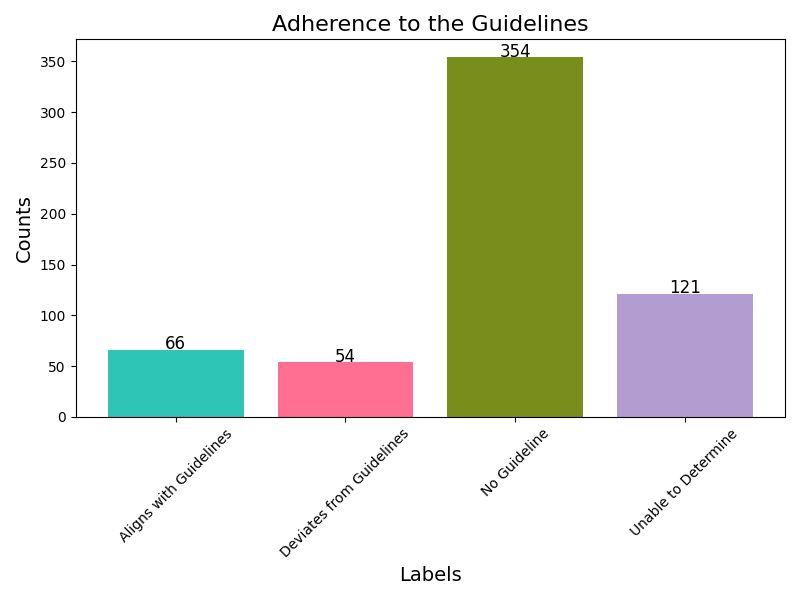}}
  \subfigure[Distribution of guideline alignment for LLM-responses \label{guideline-llms}]{\includegraphics[width=0.5\linewidth]{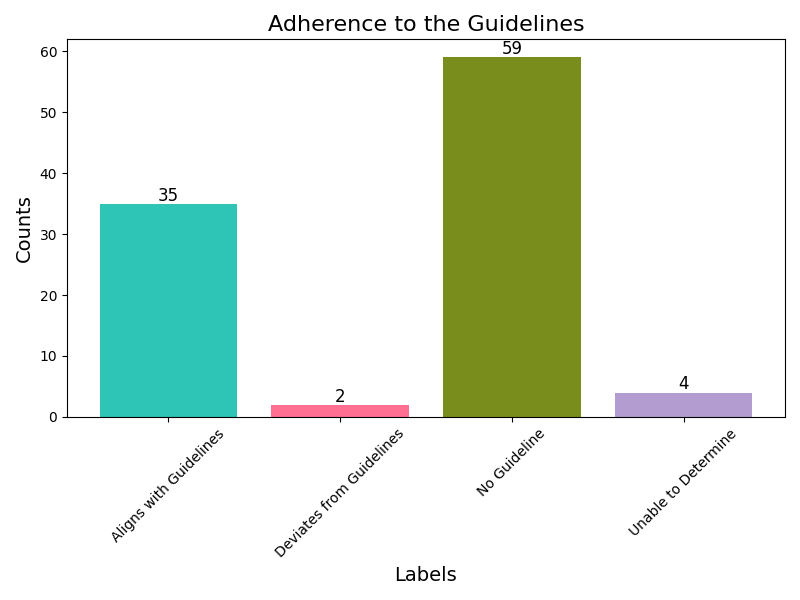}}

 \caption{Statistics on the adherence to the guideline for posts, comments, and LLM-generated responses. 'Aligns with guidelines': if the post aligns with a guideline instruction, 'Deviates from guidelines': if the post contradicts with a guideline instruction, 'No guideline': The content of the post is not verifiable with the available guideline, and 'Unable to determine': There is nothing to be compared with the guideline}
 \label{guideline}
\end{figure}

Although we originally considered four labels while checking the post contents to determine their adherence to the guidelines, we did not find any confusing or irrelevant content. So, we are not reporting the 'Unable to determine' label as part of the obtained result. The definitions and examples of the other three labels are described below. Figure~\ref{guideline-posts} presents the corresponding statistics. 

\begin{itemize}
    \item \textbf{Aligns with guidelines:} The contents of the posts that are in harmony with the guideline instructions fall under this category. Following is an example.
    
    \begin{quote}
    \textbf{EP11:} \textit{After my pregnancy test came back positive, my doctor chose to switch me to Subutex.}
    \end{quote}

    \item \textbf{Deviates from guidelines:} Along with information-seeking questions, post authors often describe their conditions and the actions they have taken. A post falls under this category if the shared actions deviate from the guidelines. The step taken by the following post author deviates from the guidelines.
    
    \begin{quote}
      \textbf{EP12:} \textit{I was experiencing severe tooth pain, which is unusual for me, and I observed that my toothache would intensify significantly after each dose. As a result, I stopped taking my dosage.}
    \end{quote}

    On the other hand, the author of the following post wants to know whether they should swallow Suboxone tablets, where swallowing is not recommended in the guidelines. 
    \begin{quote}
        \textbf{EP13:} \textit{Here are my questions: What is the initial dose for Suboxone tablets? How frequently should it be taken? How should it be ingested—should I swallow it?}
    \end{quote}

    An interest in taking opioids during recovery, as mentioned in the following post, deviates from the guidelines.
    \begin{quote}
       \textbf{EP14:}  \textit{If I run out of Suboxone two days early, can I switch to a stronger opioid for those two days and resume Suboxone immediately once I get a refill, or will I need to wait 24 hours before taking Suboxone again, as I had to when I first started it about a year ago?}
    \end{quote}
 
    \item \textbf{No guideline:} The list of authentic sources and curated guideline instructions was not exhaustive. The post authors shared many actions and statements that could not be verified due to the absence of relevant guidelines. These scenarios were captured by this category. The following is a case that we could not verify due to a lack of clear guidelines for swallowing or spitting saliva generated from the dissolved medication.
    
    \begin{quote}
      \textbf{EP15:} \textit{I recently started taking the Mylan generics and had my first evening dose last night. I've noticed that there's quite a bit of residual saliva after it dissolves. Should I swallow it or spit it out?}
    \end{quote}
    In another post, the post author asks about the particular consequences of using a substance (codeine)
    \begin{quote}
    \textbf{EP16:} \textit{If I use codeine for buprenorphine withdrawals, will I experience codeine withdrawal symptoms after just two days?}
    \end{quote}
    Another post author considered the 'spit method' as a way around headaches, an effect of medication.
    \begin{quote}
       \textbf{EP17:}  \textit{I've been learning about the spit method as a way to avoid side effects from Naloxone, like headaches or vision problems.}
    \end{quote}
    
    In the following example, the post author compares the Suboxone tablets and films with respect to the duration of relief.
    \begin{quote}
        \textbf{EP18:} \textit{Has anyone else observed that Suboxone tablets provide about one-third less duration of relief compared to the films?}
    \end{quote}

    No guidelines were available for all these cases.

    
    
\end{itemize}


In our analysis, we found that the theme "Medication during pregnancy" shows stronger alignment with the guideline, whereas posts with "Changing dosage" and "Tapering" deviate from the guidelines most, and posts with "Medication before surgery" and "Co-occurring substance usage" very often do not have any guidelines.

%% file: Tables/Stat.tex
\begin{table}[ht]
\centering
\begin{tabular}{p{3.5cm}|p{3cm}|p{3.5cm}|p{3cm}}
\hline
\textbf{Theme} & \textbf{\#of posts} &  \textbf{\#of unique post authors} & \textbf{\#of Comments}\\ \hline 
Relevant Posts & 1,951 & 1,425 & 37,276 \\ \hline 
Treatment aspects analysis & 1,000 & 812 & 18,556 \\ \hline 
Characteristics of knowledge construction analysis & 100 & 92 & 595 \\ \hline 
\end{tabular}
\caption{Table showing the number of posts, number of unique post authors, and number of comments at each step of our analysis.}
\label{tab:stat}
\end{table}

%% file: Tables/Theme_subtheme_issues.tex
\begin{table}[ht]
\centering
\begin{tabular}[!t]{p{3cm}|p{11cm}}
\hline
\textbf{Themes} & \textbf{Subthemes} 
\\ \hline



Method of administration 
& 
\begin {enumerate*} 
    \item How to dissolve, 
    \item How long it takes to dissolve, 
    \item Speculation between dissolution and a physical effect,
    \item Whether to swallow dissolved medication or not,
    \item How to swallow,
    \item Effect of swallowing,
    \item Whether to spit saliva or not,
    \item How to spit, and
    \item Effect of spitting
\end {enumerate*} 
\\ \hline

Comparing different medication options & 
\begin {enumerate*} 
\item Switching due to clinical issue,
\item Already switched vs intent to switch,
\item Switching due to logistical / systematic barrier, and
\item Combining multiple medication options
\end {enumerate*} 
\\ \hline

Psychophysical effects associated with medication
&
\begin {enumerate*} 
\item Diversity of perceived effects,
\item How others address these effects,
\item Severity, frequency, duration of effects, and
\item Perceived trigger of effect 
\end {enumerate*}  
\\ \hline



Tapering 
&
\begin {enumerate*} 
\item Tapering method,
\item Tapering goal,
\item Tapering timeline,
\item Effects of tapering and how to cope with them, and
\item Reason of tapering 
\end {enumerate*} 
\\ \hline

Co-occurring alcohol consumption
&
\begin {enumerate*} 
\item Frequency and degree of alcohol,
\item Reason of usage,
\item Effect of consumption, and
\item Polysubstance use
\end {enumerate*} 
\\ \hline

Co-occurring substance usage
&
\begin {enumerate*} 
\item Type and amount of substance usage,
\item Reason of usage,
\item Effect of substance usage, and
\item Polysubstance use
\end {enumerate*} 
\\ \hline

Medication during pregnancy
&
\begin {enumerate*} 
\item Tapering during Pregnancy,
\item Comparing effective option,
\item Managing withdrawal symptoms during pregnancy, and
\item Consulting doctors
\end {enumerate*} 
\\ \hline

Medication before surgery
&
\begin {enumerate*} 
\item Concern about medication changes during surgery, and
\item Query about dosing during surgery
\end {enumerate*} 
\\ \hline

Resuming treatment after return to usage
&
\begin {enumerate*} 
\item Waiting time after the last substance use,
\item Starting dosage,
\item Medication option to start treatment, and
\item Communicating with provider about resuming treatment for safety
\end {enumerate*} 

\\ \hline


\end{tabular}
\caption{List of 9 themes and 40 sub-themes emerged through our analysis, highlighting the complexity of a socially constructed treatment plan for Suboxone, a medication for opioid use disorder (MOUD).}
\label{theme_subtheme_issues}

\end{table}

%% file: Sections/5_Peer_Suggested_Strategies.tex
\section{Strategies emerging from peers' comments}
\label{peer-suggestions-section}

To gain deeper insights into socially constructed treatment plans, we conduct an in-depth analysis of peer responses. We mainly focused on two sub-goals- 1) the characteristics of knowledge construction by analyzing the comments from the peers against the queries or concerns expressed by the post authors, and 2) whether the suggestions or advice made by the peers follow the authentic guidelines.

\subsection{Methods}
\subsubsection{Identifying characteristics of knowledge construction}
\label{peer-suggested-strategies}
We first filtered out the peer-generated responses against the 100 posts obtained in Section~\ref{aspects_methods}. To simplify the following annotation, we considered all the comments from a post if the number of comments is less than or equal to 10. If the count is more than 10, we considered the first ten comments sorted by the ratings provided by Reddit. Thus, we obtained 595 comments in total. The first two authors had the contexts from the posts and also explored some of the comments to know about the data. The authors then performed a reflexive thematic analysis~\cite{braun2006using} on those comments, keeping the experts in the loop. However, in this case, our target was to determine different strategies that peers employ while commenting on the posts. 

\subsubsection{Aligning peer-provided comments to the guidelines}
\label{comment-adherence-method}
Employing a qualitative approach, we compared each peer-generated response with the relevant authoritative guidelines. We used the guidelines enlisted in Section~\ref{aspects_methods} as the references for the evaluation. We annotated each response with one of the four labels: ``Aligns with guidelines," ``Deviates from guidelines," ``No guideline," and ``Unable to determine." We followed the same strategy of annotation as performed in the last section, i.e., annotation by the first author, followed by the review of the second author, and resolution by discussion. We also used the same sources for fact-checking with clinical guidelines listed in Section \ref{subsubsec:method_aspects}.

\subsection{Results}

From our analysis, we found five different strategies used by the peers while responding to the post authors. These strategies resemble the characteristics of constructed knowledge in the communities. We also assessed the reliability of the peer-generated comments by analyzing the statistics of their guideline-following tendency.

\subsubsection{Characteristics of knowledge construction:} 
\label{comment-strategies-result}
Our analysis revealed the five distinct strategies that peers employ to construct knowledge by commenting to the information-seeking ports. Below, we define and provide examples of these strategies:

\begin{itemize}
    \item \textbf{Personal treatment experience}: This strategy involves individuals sharing their personal experiences with treatment. For example, a user was asking ``\textit{how long to wait after placing the pill under the tongue?}". The peer responded with the following, 
    \begin{quote}
        \textbf{EC1:} 
        \textit{Avoid holding excess saliva in your mouth, as it can hinder absorption. If you haven’t tried it, crush the pill into powder and place it under your tongue, pressing down with your tongue to minimize saliva. Administering half the dose at a time might also help. Excess saliva can cause the pill to move around or get swallowed, reducing absorption...}
    \end{quote}
    
Although this was not mentioned in any of the guidelines, our clinical experts verified this information. They also verified the following suggestion from another peer in the same context: "leaning forward or looking down while taking the medication may help."
    
    \item \textbf{Personal lived experience}: Due to the anonymity of Reddit, people freely share their personal life experiences while offering suggestions. This includes insights into their daily lives, family dynamics, and overall quality of life. For instance, one peer shared how this medication drastically improved their quality of life to encourage others not to taper the medication. Tapering is a common self-management strategy that is often linked with stigma towards long-term treatment~\cite{dickson2022you}. 

    \begin{quote}
        \textbf{EC2:} \textit{I'm 47 years old and have been on buprenorphine for about twenty years now.... Never abused them and haven't had a relapse or a drink of alcohol since. Lots, I mean lots of my friends that were on subs and got off them, are unfortunately all dead and left their families behind so I'm a lifer and I've already accepted it. Quality of life in this short trip on this rock floating through space! Good luck everyone!}
    \end{quote}
    Here, "lifer" is a term commonly used in MOUD online communities to refer to someone committed to long-term MOUD treatment for recovery. This comment also highlights how peers negotiate the meaning of the treatment process and goals. 
    
    \item \textbf{Clinical strategy:} Peers often suggest various clinical strategies through their comments, either directly or indirectly recommending clinical options or solutions. For example, one asked about the strategy of tapering. A peer suggested the following strategy to taper the medication slowly: 
    \begin{quote}
        \textbf{EC3:} \textit{Do not taper too fast. If you are in 8mg, the best way is to go from 8 to 4.  Then to 3, 2, 1.5, 1, .75, .5, .25.  Then start skipping days. You will go through hell if u do rapid detox. Please trust me.}
    \end{quote}
    There are no clinical guidelines for tapering. However, our clinical experts deem this to be a valid suggestion. They also highlight the need to customize the tapering plan based on how patients respond to the plan.
    
    \item \textbf{Psychological strategy: }This strategy involves peers offering moral support, motivation, and psychological advice. For instance, in response to someone's concern about their medication not being absorbed properly, a peer shares situated knowledge and highlights how women might have more challenging treatment experiences due to hormonal fluctuations as a coping strategy.
    
    \begin{quote}
        \textbf{EC4:} \textit{Several factors can affect how you feel, including your sleep, metabolism, vitamin and mineral levels, hydration, and hormone levels (particularly for women). For instance, I need up to double the usual dose during PMS to feel okay. The physiological factors and uneven distribution of buprenorphine on the strips can make consistent dosing challenging, and this issue seems more common in women due to hormonal fluctuations.}




    \end{quote}
    
    \item \textbf{Speculation:} At times, peers speculate while providing suggestions based on the information shared by the original post authors. For example, a post author mentions being able to stay away from all opiates without any treatment and asks whether they should start Suboxone. In response, a peer speculates the original poster (OP) might not need any treatment at all unless they (OP) feel they might return to substance use. 

    \begin{quote}
        \textbf{EC5:} \textit{If you've already been off everything for 50+ days, I'm not sure Suboxone is the answer. However, it might be the lesser of two evils if you're at risk of relapsing.}
    \end{quote}
   
\end{itemize}
\begin{figure}[h!]
  \centering
  \subfigure[Statistics of strategies discussed by the peers]
  {\includegraphics[width=0.45\linewidth]{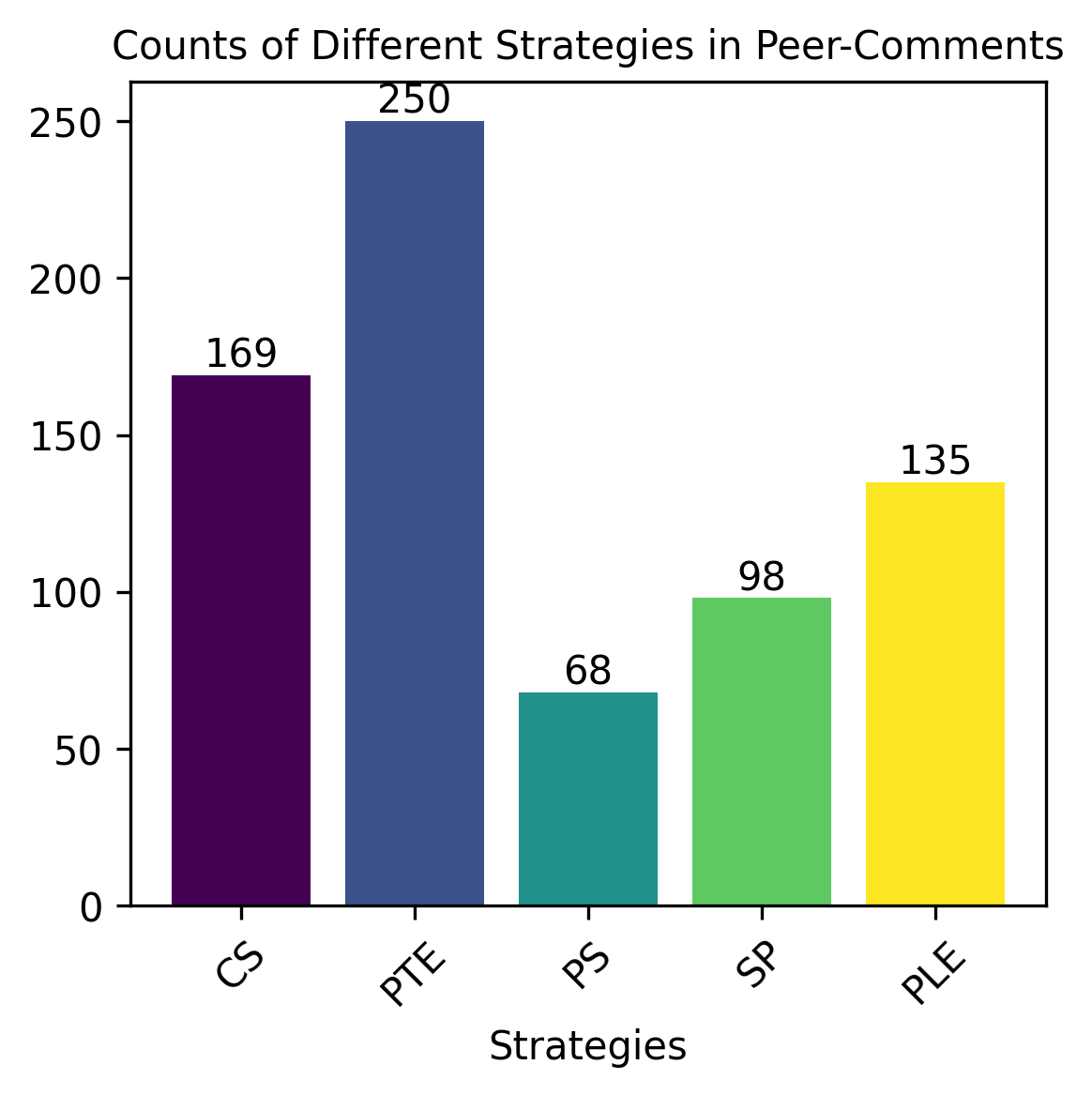}}
  \subfigure[Comparison of statistics of GPT-4 and Peer-Comments]{\includegraphics[width=0.49\linewidth]{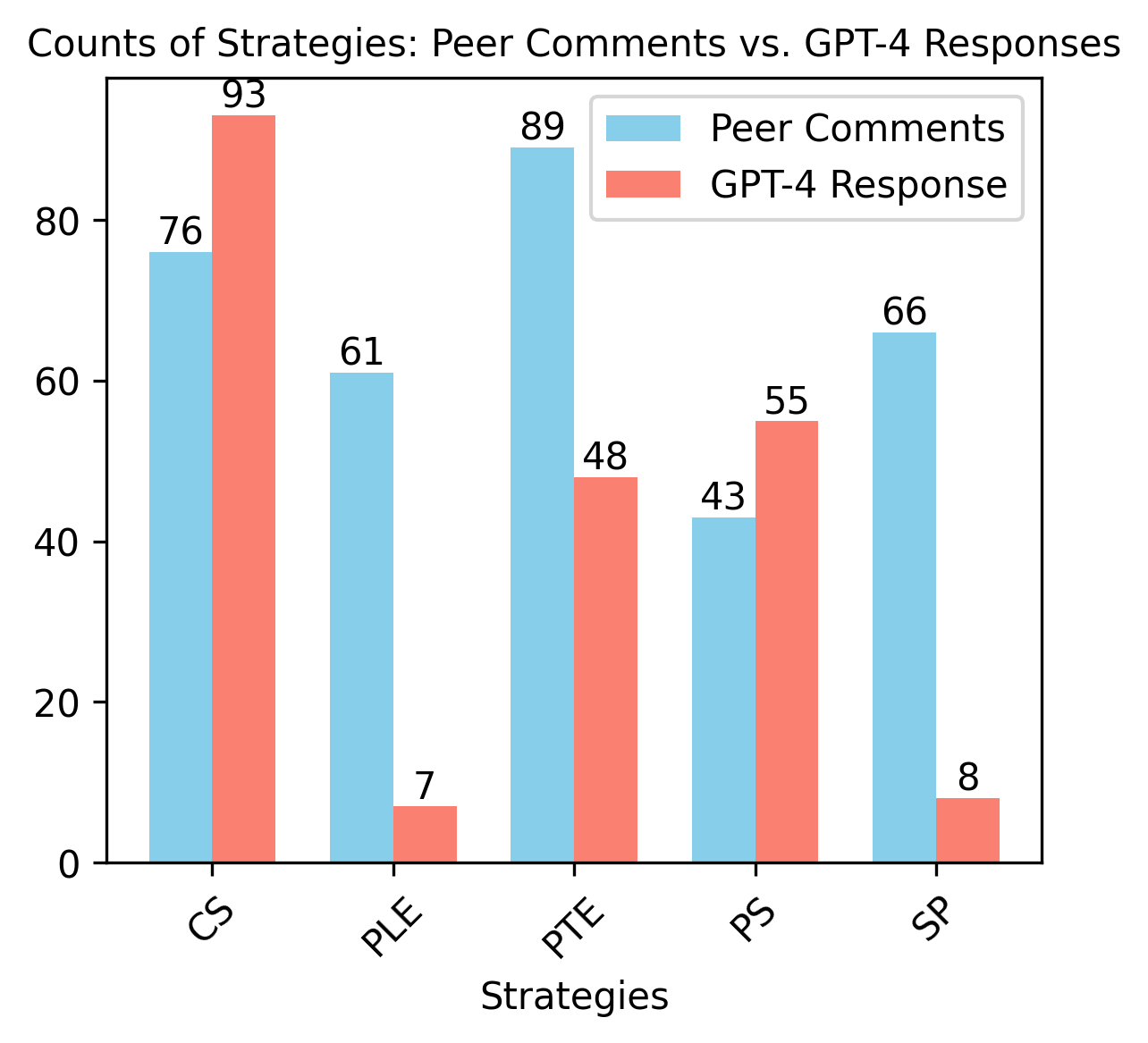}}

 \caption{Statistics of different strategies discussed by peers and GPT-4. Here, CS, PTE, PS, SP, and PLE indicate `Clinical Strategy', `Personal Treatment Strategy', `Psychological Strategy', `Speculation', and `Personal Lived Experience' classes, respectively. Peer comment counts are normalized while comparing with GPT-4 response count to avoid counting the same strategy twice. We took the set of strategies that were discussed in the comments.} 
 \label{stratagies-statistics}
\end{figure}

The distribution of the strategies is presented in Figure~\ref{stratagies-statistics}(a). Our analysis shows that "Personal Treatment Experience" is the most common strategy employed by the peers, followed by "Clinical Strategy." On the other hand, "Psychological Strategy" was the least available in our study.


\subsubsection{Adherence to the Guideline:}
The result of our analysis comparing peer responses with clinical guidelines is presented here. We applied four labels for categorization: "Aligns with guideline," "Deviates from guideline," "No guideline," and "Unable to determine." Below, we provide the definitions and examples for each label, with the corresponding distribution illustrated in Figure ~\ref{guideline-comments}.

\begin{itemize}
    \item \textbf{Aligns with guidelines:} The peers' responses that follow the guidelines are marked with this label. The following response is from a peer for a post where the post author expressed her concern about Suboxone usage during pregnancy. The response suggesting trying another mediction option (i.e., Subutex) follows the guideline instructions.
    
    \begin{quote}
      \textbf{EC6:} \textit{Seek a different doctor who will prescribe Subutex. Naloxone in Suboxone can be harmful to a fetus, and the stress from a quick taper could negatively impact your unborn baby.}
    \end{quote} 

    \item \textbf{Deviates from guidelines:} Peers provide diverse comments in response to posts, and many of these can deviate from clinical guidelines. Such comments are categorized as``Deviates from guidelines." For instance, despite guidelines advising against abruptly stopping medication, one peer suggested the following:
    

    \begin{quote}
        \textbf{EC7:}  \textit{I went to rehab and stopped [medication] after taking 4 mg every other day.}
    \end{quote}
    
    

    Although snorting is clinically discouraged, the peer suggests the following in a comment in another post about snorting medication.

    \begin{quote}
        \textbf{EC8:} \textit{I personally snort the medication and find it more effective than taking it sublingually. It's safe to use it this way now.}
    \end{quote}

    For a question where the post author asked whether they can mix alcohol with medication, a peer provided the following comment that contradicts the guidelines.
    \begin{quote}
        \textbf{EC9:} \textit{There's no issue with drinking alcohol while on medication; combining it with medication is not harmful.}
    \end{quote}

    \item \textbf{No guideline:} Some responses could not be verified, as we could not find any guidelines for them. These responses were labeled as "No guideline." We could not verify the following response as there was no direct evidence in the guidelines mentioning the relation between swallowing and headaches.

    \begin{quote}
        \textbf{EC10:} \textit{Swallowing can lead to severe headaches.}
    \end{quote}
    

    The following comment from a peer on the consequences of using codeine for buprenorphine withdrawals could not be verified in the guidelines.
    \begin{quote}
        \textbf{EC11:} \textit{It's not advisable at all. Codeine is a full agonist, while buprenorphine is a partial agonist. Using codeine may not fully alleviate withdrawal symptoms, but it could undermine the healing benefits you've gained from being on a partial agonist.}
    \end{quote}


    In a post where the post author seeks suggestions about which medication form (strip vs. tablet)is better, peers offer perspectives on both options. There is no clinical evidence on which form is better, and it depends on individual preference. Our clinical experts also confirmed this. This also illustrates potential sources of contested knowledge in socially constructed treatment plans.
    
    \begin{quote}
       \textbf{EC12:} \textit{Films wrecked my gums and tongue, which is why I switched to tablets. Otherwise, I would have stayed with them because they dissolve faster at higher doses than the tablets.}
    \end{quote}

    \begin{quote}
        \textbf{EC13:} \textit{I prefer films any day. They're much easier to cut for dosing, and I find the tablets taste awful compared to the films.} 
    \end{quote}

      \item \textbf{Unable to determine:} There were some comments that were ambiguous or not relevant, and we were unable to ground them to clinical guidelines. Following is an example comment where the peer suggested checking another subreddit for the issue discussed in the post.
    
    \begin{quote}
      \textbf{EC14:} \textit{Visit the Methadone subreddit to see that they're experiencing similar issues as us. It's worth checking out.}
    \end{quote}
\end{itemize}

Our analysis revealed that comments on posts related to the themes of "Medication during pregnancy" and "Co-occurring substance usage" were more likely to align with the guidelines. Conversely, comments on posts discussing ``Changing dosage" were more prone to contradict the guidelines. ``Medication before surgery," ``Comparing different medication options," and ``Resuming treatment after returning to usage" contained more comments that fell under the "No guideline" category.


%% file: Sections/6_LLMs_Responses.tex
\section{Can the LLMs reflect the socially constructed knowledge}

In this section, we aim to investigate the ability of LLMs to capture and reflect socially constructed treatment knowledge. We conducted two experiments: 1) analyzing the distribution of characteristics within the constructed knowledge, specifically focusing on the types of strategies present in LLM-generated responses, and 2) comparing these responses against the established authentic guidelines.

\subsection{Methods}

\subsubsection{Reflection of the treatment strategies in LLMs}
For our analysis of LLMs, we selected ChatGPT as the representative model. This choice was based on its widespread popularity and its accessibility as an interactive chatbot, making it easily usable by the general public. We used GPT-4o API in this experiment \cite{openai2024gpt4technicalreport}. We provided the LLM with an information-seeking post and used a prompt to direct its response for that post \cite{min-etal-2022-rethinking}. The following prompt was used in our analysis.

\begin{quote}
    \textit{You are engaging in a conversation with a human. This is someone asking question on opioid use disorder and its treatment on a MOUD-related subreddit. Respond to the following question using on average 50 words and a maximum of 150 words. Your response will be visible to anyone visiting the subreddit.}
\end{quote}

We used an average of 50 words and a maximum of 150 words to restrict the output of the LLM-generated response. The numbers used here are close to the average and 90\% percentile of the lengths of the previously used peer-provided comments, respectively.

Using LLM, we generated responses for the 100 posts used in~\ref{post-adherence-method}. The second author labeled each LLM-generated response with one or multiple strategies from a set of strategies identified in~\ref{comment-strategies-result}. The first author then reviewed and resolved the disagreement by discussions with the second author. Moreover, we wanted to compare the prevalence of different strategies in the peer-provided comments and LLM-generated responses. Remember that we used the comments of the same posts in Section \ref{peer-suggestions-section} to identify the strategies adopted by the peers while responding to a post. 

\subsubsection{Adherence to the Guideline:}
Similar to Sections~\ref{post-adherence-method} and~\ref{comment-adherence-method}, we performed an annotation to check whether the LLM-generated response adhere to the guidelines. The second author first labeled each response, followed by a review of the first author and a resolution by discussions.

\subsection{Results}
Here, we first present our result of the comparison between treatment strategies available in comments and LLM-generated responses. We then report our findings on the guideline-following tendency of the responses by LLM.

\subsubsection{Reflection of the treatment strategies in LLMs:}
Our analysis revealed the frequency of socially constructed treatment strategies in LLMs. We can see LLMs offer ``Clinical strategy" most of the time. The next strategy employed by it is ``Psychological Strategy."  Figure~\ref{stratagies-statistics}(b) shows the comparative distribution of strategies in peers' comments and LLM responses. This comparison clearly shows that LLM could not capture the strategies ``Personal Lived Experience" and ``Speculation," which are commonly available in LLM-generated responses.

\subsubsection{Adherence to the Guidelines:}
 In this section, we present the result of our guideline adherence analysis on LLM-generated results. Similar to the comments, we classified the responses into four categories as described below. The relevant distribution can be found in Figure~\ref{guideline-llms}.

\begin{itemize}
    \item \textbf{Aligns with guidelines:} The LLM-generated responses that follow the guidelines are marked with this label. The following is a response provided by the LLMs against the question of whether to use codeine for buprenorphine withdrawal.
    
    \begin{quote}
      \textbf{LR1:} \textit{Using codeine to relieve buprenorphine withdrawal can offer short-term relief, but it may cause its own withdrawal symptoms after only a few days. As an opioid, even brief use of codeine can lead to dependence.}
    \end{quote} 

    The following response to another question about how long to wait after fentanyl use aligned with the guidelines.

    \begin{quote}
       \textbf{LR2:} \textit{Generally, after using fentanyl, it's recommended to wait at least 24-72 hours.}
    \end{quote}

    \item \textbf{Deviates from guidelines:} If the response of the LLMs conflicts with or contradicts the guidelines, we marked it with ``Deviates from guidelines". For example, the following guideline-violating suggestion (i.e., tablets are swallowed) was provided by the LLM when one post author asked about the difference between tablets and films.
    
    \begin{quote}
    \textbf{LR3:} \textit{The primary difference between buprenorphine tablets and films is their formulation and how they are administered. Tablets are swallowed, whereas films dissolve under the tongue, enabling faster absorption.}
    \end{quote}
 
    \item \textbf{No guideline:} LLMs also provided some responses that we could not verify as our list of guidelines was not exhaustive. These responses were labeled as ``No guideline." For instance, in response to a question about whether to use a sublocade shot while stopping Suboxone, the LLM generated an answer that falls under this label.
    
    \begin{quote}
      \textbf{LR4:} \textit{Sublocade is administered as a depot injection, allowing buprenorphine to be released gradually over time. This slow release helps prevent withdrawal and cravings while avoiding a significant high.}
    \end{quote}
    In response to the question on how long to use medication after two weeks of oxycodone use, the following answer was provided by the LLM, which we were unable to verify.

    \begin{quote}
       \textbf{LR5:} \textit{After a short period of heavy oxycodone use, starting Suboxone can help manage withdrawal and cravings. Typically, it's recommended to stay on it for at least a few weeks to stabilize.}
    \end{quote}

    For the question where the post author reported headache as an effect of Suboxone, LLM provided the following response, for which were no relevant guideline found.
    \begin{quote}
      \textbf{LR6:} \textit{It’s possible that your headaches could be related to Suboxone, especially since you've switched back recently. Some people report headaches as a side effect, and the timing of your dose might also play a role. }
    \end{quote}

      \item \textbf{Unable to determine:} The ambiguous or irrelevant responses of the LLMs were categorized as ``Unable to determine". For example, the following response was ambiguous as breaking the tablets is not aligned with the guidelines. At the same time, it also suggested following the manufacturer's guidelines.
    
    \begin{quote}
     \textbf{LR7:} \textit{For sublingual tablets, breaking them into smaller pieces or powdering them might improve absorption. However, it's crucial to adhere to the manufacturer's guidelines, as some tablets are formulated for specific release profiles and should not be cut into smaller pieces.}
    \end{quote}
\end{itemize}

LLMs provide better-quality and guideline-following responses for ``Medication during pregnancy" and ``Tapering"-related questions. We found only two pieces of evidence where LLM-provided responses deviate from the guidelines - one in ``Comparing different medication options" and another one in ``Resuming treatment after returning to the usage." ``Co-occurring substance usage" and ``Psychophysical effects associated with medication" were the themes where LLMs provided responses that were not verifiable by the guidelines.

%% file: Sections/7_Interviews.tex
\section{Interviews}
Along with expert feedback throughout our study, we conducted semi-structured interviews with three clinicians. The size of the interview participants was small as our purpose was to incorporate unbiased feedback from the clinicians who were not part of the author list or the expert team associated with our study. The method and results of the interviews are presented in this section

\subsection{Method}
We conducted semi-structured interviews with three clinicians to collect feedback on a set of posts and associated comments and LLM responses. We selected 23 posts covering each of the themes and prevalent subthemes. Each interview lasted about an hour. We are referring to the clinicians as P1, P2, and P3. P1 and P2 were male clinicians with more than twenty years of experience, while P3 was a female clinician with more than fifteen years of experience. 

\subsection{Results}
Some of the key observations found in our analysis are highlighted here.

\textbf{Situated and contested knowledge around the practice of taking medication}: Clinicians highlight how certain aspects of the treatment plan have contested knowledge. Such as suggestions not to crush pills and cut strips are clinically valid guidelines as they reduce the risks of diversion. However, if someone crushes the pill, it can still work as long as they take the intended dose. They also acknowledge that patients might crush the pill or cut their strip for better dissolution while reducing the dose during tapering. 
\begin{quote}
    \textit{"Officially, it's not recommended to divide the film because the medication might not be evenly distributed, which can lead to inconsistent dosing—one half might have more medication than the other. However, sometimes dividing the film is necessary, especially when the required dose isn't available in a single formulation. For example, if someone is tapering down to one milligram, but the smallest available dose is two milligrams, they may need to divide the film. Despite the official advice, many people still divide films and tabs and find it works well enough." [P1]}
\end{quote}

\textbf{Effectiveness of peer suggested strategies:} Clinicians highlight certain peer strategies that are not in the guideline but still could be helpful. Such as all participants agreed that the following peer-suggested strategies could be helpful with the absorption of the medication and avoid potential concerns about dissolution. 
\begin{quote}
    \textit{"Leaning forward or looking down while taking the medication may help absorption"[P2]}
\end{quote}
\begin{quote}
\textit{"Using a small amount of alcohol or brushing teeth before taking bupe may aid in dissolution."[P3]}
\end{quote}

They also highlight that in certain settings, like prisons or hospitals, medication is often crushed before administration. This is done to prevent misuse or diversion of the medication. 
\begin{quote}
    \textit{"In many cases, especially in incarcerated settings or hospitals, the medication is crushed before being given to reduce the risk of diversion. Even when crushed, the medication still provides the full dose needed." [P1]}
\end{quote}

Clinicians also pointed out several strategies and suggestions from peers that are not in the clinical guidelines but still could be reasonable. For example, one clinician pointed out that in some clinical guidelines of Suboxone, "headache" is listed as a potential side effect, but it does not capture how severe it can be, as often found in the narrated lived experience. 
\begin{quote}
    \textit{"It [clinical guideline] minimizes the potential difficulties of headaches and how debilitating it can be too. So  I think, like each of these responses [LLM, clinical guideline, and peer responses] have some truth and maybe some things that they're missing." [P1]}
\end{quote}

Thus, clinical guidelines and self-reported lived experience can complement each other.

\textbf{Limitations of socially constructed plan:}  Clinicians pointed out while some peer strategies are helpful, some can be potentially harmful, e.g. "taking psychedelics to treat headache," "taking kratom to taper Suboxone". They also pointed out certain strategies are impractical, e.g. "waiting 6o minutes for the medication to dissolve under the tongue while sitting still."

Their perspectives also varied from patients in OHCs. For example, while many patients shared preferences for one form of medication over another (i.e., pills and strips), one of our participating clinicians acknowledged how personal preferences might vary. He also pointed out:

\begin{quote}
    \textit{There should not be a significant difference between strips and pills ... but it is unlikely that it would lead to a clinically detectable change in outcome. [P2]}
\end{quote}

They also pointed out that reading about adverse psychological and physical effects of treatment might cause the "nocebo effect." The nocebo effect is when a patient's negative expectations about a treatment cause a worse outcome than would otherwise have occurred. 

Another observation was when trying to respond to these questions asked by patients, clinicians often ask several followup questions to understand the context and resolve any ambiguity. This strategy is totally missing from LLM-generated responses and seldom reflected in peer responses. For instance, how to resume treatment after a return to using opioids is a prevalent aspect of a socially constructed plan. There are not a lot of details in the clinical guidelines. Clinicians participating in our study pointed out how they ask a series of questions to determine treatment plan, including, what the process has been like for the patient in the past, their prior induction process and how it went. They also want to understand how much the patient already knows and their preferences, e.g.,  how long they are willing to wait. They also want to learn about the patient's home environment, responsibility, co-occurring diagnosis, and severity of substance use disorder to facilitate a shared decision-making process.


%% file: Sections/8_Discussion.tex
\section{Discussion}
Our results highlight that the constructed knowledge in online communities with uncertain and complex medical conditions, like MOUD, is very complicated in nature, as evident by multiple themes, sub-themes, and strategies obtained from our analysis. We also find the prevalence of online content with no guidelines or deviates from the guidelines. The lack of proper guidelines can create confusion and lay the ground for rumors. We also demonstrate that LLMs are better at following guideline instructions but lack some important human properties, suggesting a complementary use of the LLMs as a technology-mediated tool for information support, along with online communities. Overall, our study uncovers several fundamental research gaps in the knowledge construction process in online communities with complex medication conditions. These gaps need to be addressed by the HCI community.




\begin{itemize}


\item \textbf{Impact of socially constructed complex knowledge about treatment plans:} Online communities create complex knowledge about treatment plans, as evident by nine themes and forty sub-themes regarding treatment aspects and five different strategies generated by interactions among community members. This diversity is captured from 1,000 discourses associated with a single type of discussion classified as 'Taking MOUD.' Increasing the sample counts and considering data from other categories would provide us with a more complex set of treatment plans. This complexity can affect the community members in many ways. 

Due to the availability of discussions in diverse aspects, patients can find out the discussions related to their problems and connect with their peers in the same situation, offering situated knowledge~\cite{lee2014effects}. Situated knowledge is particularly important for people with complex medical conditions. Being inspired by peers in the same boat, people can freely share their concerns and problems without the fear of being judged. For example, pregnancy while on MOUD treatment is considered stigmatized, and a lot of women do not want to inform others about the treatment for fear of losing custody of the child. However, they feel free to share their concerns in the communities, like the question asked in \textbf{EP9}, as they know there are other women who are going or have been in the same situation.

The complexity of knowledge can affect community members negatively, too. Peers often share their ``Personal treatment experience," which might not be clinically verified and can even be wrong. For example, ``crushing the pill into powder" as suggested by peer in \textbf{EC1} is not clinically verified. Following these types of suggestions can lead to harmful treatment outcomes. Moreover, the prevalence of contested knowledge, where two different peers provide completely opposite suggestions, might confuse the patients or cause the "Nocebo effect"~\cite{colloca2011nocebo}. 


The dual nature of this socially constructed complex treatment plan needs more scrutiny. The dependence of the vulnerable group of people on online communities demands that HCI researchers explore this space more to reduce harm while ensuring a comfortable space for those people.

\item \textbf{Tension between hermeneutics and heuristics:}
The heuristical approach~\cite{metzger2010social} has always been hailed in online communities because individuals come to social media to provide fast and easy access to information. Heuristics emphasizes quick and practical solutions by simplifying complex problems. While seeking information, people often prioritize getting quick solutions. In online health communities, information exchange and knowledge propagation are very complex. While annotating data, we found one discourse can contain discussions related to several themes. It indicates the issues are interdependent. For example, the post author of \textbf{EP4} talks about headache (theme ``Psychophysical effects associated with medication") as well as the evidence of swallowing (theme ``Method of administration"). That's why people surfing sensitive online health communities like MOUD, where knowledge creation is complex, need a deeper, thorough, and hermeneutical interpretation of content and context. HCI researchers need to focus on the online health communities to balance hermeneutics and heuristics. Research is required to ensure the hermeneutical requirement of 'scientific evidence and rationale' while maintaining the heuristical benefit of 'capturing lived experiences.'


\item \textbf{Rumors during knowledge construction:}
Although rumors are prevalent in online communities~\cite{spiro2023rumors}, these are very hard to define and identify in online health communities. Knowledge construction and sensemaking processes in complex, uncertain topics are susceptible to rumors. Our study presents that there are lots of issues for which no relevant guidelines are available. For example, clinical guidelines instruct that Suboxone film should not be swallowed; rather, the patients should let it dissolve under the tongue. However, no relevant guideline is available describing what to do with the saliva that is generated after the dissolution of the medication as mentioned by the post author of \textbf{EP15.} The lack of proper guidelines provides the breeding ground for the rumors. People relying on information in online communities can easily be impacted by the rumors. Rigorous research is required to explore how to define these rumors and when, why, and how rumors are created and spread. HCI community should focus on this area since rumors can make the certain people more vulnerable.

Emphasizing "authenticity over authority" can play a crucial role in addressing rumors. Peer-support technology can leverage this principle to foster a more genuine exchange of information. By prioritizing real experiences and authentic voices over traditional authoritative sources, these platforms can help debunk harmful rumors, reduce stigma, and combat misinformation effectively. This approach is required to create a safer and more informed community environment.

\item \textbf{Social construction of meaning:} Language plays a crucial role in social constructionism. It enables individuals to communicate and negotiate meanings with one another. The interpretation of language can differ between patients and providers. For example, there is confusion about "dissolving" the medication. Guidelines suggest letting the medication dissolve under the tongue for 10 minutes, but there are no guidelines on what to do with the saliva produced during dissolution. This confusion has led to terms like "spit," "swallow," and "wait time after dissolution." Post \textbf{EP1} illustrates such confusion about the term "dissolving." Another example is  "tapering", a long-term process from the clinicians' perspective spanning over at least year if not longer. However, patients often consider tapering as a rapid process and even plan for tapering at the beginning of treatment. During the tapering process, patients often "cut" the dosage to take an amount lower than the smallest available film/tablet (e.g., patients may want to take 1 mg while the smallest available film for Suboxone is 2 mg). Cutting the film has become a common discussion issue among patients, although guidelines prohibit this practice as it may reduce the medication's effectiveness. However, due to the prevalence of this behavior, clinicians have recently become aware of this fact and acknowledged that cutting the dosage is done by the patients when they are on a dosage (e.g., 1 mg) smaller than the smallest strength of available medication (e.g., 2 mg is the smallest film strength from Suboxone).

Research is required to ensure that the diverse voices of the patients are captured. Patient-centered communication, where patients experience active listening without judgment, can help capture the premise of the language. On the other hand, patient-provider communication, which is often dominated by the providers, needs significant research so that patients can get the space to share their constructed and interpreted meaning with the providers. All these are open research questions that need to be addressed.

\item \textbf{The role of technology-mediated tools in complex medical conditions:} People come to online communities with complex medical conditions to fulfill both their informational and emotional needs. We show that the knowledge generated by humans covers both informational and emotional needs, but the participants' subjective experiences often dominate the quality of information. On the other hand, technology-mediated solutions can offer more objective and reliable information. Our exploration of the LLMs, one of the most promising technological solutions for informational support, demonstrates that LLMs can provide more reliable and safer information. However, they often lack some crucial human aspects. We argue that technological solutions like those provided by LLMs can be complementary to existing support systems in online communities and should work as assistants to humans. How to effectively combine the ``humanness" of humans and the ``reliability" of technological solutions can be an impactful HCI research area in the context of surfacing socially constructed knowledge.

\end{itemize}

%% file: Sections/9_Limitations_and_Future_Works.tex
\section{Limitations and Future Works}
We acknowledge that our study has several limitations. A larger number of posts and comments from online communities would likely yield better results and help us identify more diverse treatment plans. In addition, using a different set of samples might generate different treatment aspects. Data from more subreddits and other communities related to the same medical condition could provide a more comprehensive view of treatment aspects. 

In the future, we plan to incorporate deeper-level comments (i.e., comments responding to other comments on a post) to capture more nuanced interactions among community members and to understand how post authors and other members participating in the discussion accept different types of suggestions. We also aim to perform a detailed exploration of the scope and limitations of LLMs in mirroring socially constructed knowledge. We used the latest version of ChatGPT/GPT-4 API in this paper due to its superior performance over other LLMs, and plan to scope open-source LLMs for this task in the future.

The goal of this paper was to characterize how the socially constructed protocol align with clinical guidelines, understand the cases where it deviates. and identify to which extent they are captured in LLMs. In the futuer we plan to extend this study by engagin with patients in these communities through focus groups and understand how they make sense of different aspects of socially constructed treatment plan.



%% file: Sections/10_Conclusion.tex
\section{Conclusion}
Our study highlights the critical role of social constructionism in healthcare, particularly through the collective sensemaking process in online communities, which shapes treatment plans for individuals with opioid use disorder and similar stigmatized, complex conditions. We demonstrate the complexity and diversity of aspects of treatment plans, identify potential knowledge gaps, propose possible solutions, and highlight research questions for future investigation. We also examine the potential and current limitations of LLMs in capturing the nuanced, socially constructed knowledge shared among peers. These findings are intended to improve overall treatment quality by aiding the patients in making better-informed decisions and assisting healthcare providers in understanding patient experiences and preferences.

%% file: MAIN_submission.bbl

\begin{thebibliography}{45}


\ifx \showCODEN    \undefined \def \showCODEN     #1{\unskip}     \fi
\ifx \showDOI      \undefined \def \showDOI       #1{#1}\fi
\ifx \showISBNx    \undefined \def \showISBNx     #1{\unskip}     \fi
\ifx \showISBNxiii \undefined \def \showISBNxiii  #1{\unskip}     \fi
\ifx \showISSN     \undefined \def \showISSN      #1{\unskip}     \fi
\ifx \showLCCN     \undefined \def \showLCCN      #1{\unskip}     \fi
\ifx \shownote     \undefined \def \shownote      #1{#1}          \fi
\ifx \showarticletitle \undefined \def \showarticletitle #1{#1}   \fi
\ifx \showURL      \undefined \def \showURL       {\relax}        \fi
\providecommand\bibfield[2]{#2}
\providecommand\bibinfo[2]{#2}
\providecommand\natexlab[1]{#1}
\providecommand\showeprint[2][]{arXiv:#2}

\bibitem[Amato et~al\mbox{.}(2011)]%
        {amato2011psychosocial}
\bibfield{author}{\bibinfo{person}{Laura Amato}, \bibinfo{person}{Silvia Minozzi}, \bibinfo{person}{Marina Davoli}, {and} \bibinfo{person}{Simona Vecchi}.} \bibinfo{year}{2011}\natexlab{}.
\newblock \showarticletitle{Psychosocial and pharmacological treatments versus pharmacological treatments for opioid detoxification}.
\newblock \bibinfo{journal}{\emph{Cochrane Database of Systematic Reviews}} \bibinfo{number}{9} (\bibinfo{year}{2011}).
\newblock


\bibitem[Andrews(2012)]%
        {andrews2012social}
\bibfield{author}{\bibinfo{person}{Tom Andrews}.} \bibinfo{year}{2012}\natexlab{}.
\newblock \showarticletitle{What is social constructionism?}
\newblock \bibinfo{journal}{\emph{Grounded theory review}} \bibinfo{volume}{11}, \bibinfo{number}{1} (\bibinfo{year}{2012}).
\newblock


\bibitem[ASAM(2024)]%
        {ASAM}
ASAM \bibinfo{year}{2024}\natexlab{}.
\newblock \bibinfo{booktitle}{\emph{American Society of Addiction Medicine (ASAM)}}.
\newblock
\urldef\tempurl%
\url{https://www.asam.org/}
\showURL{%
\tempurl}
\newblock
\shownote{Last Accessed: 2024-08-10}.


\bibitem[Balsamo et~al\mbox{.}(2023)]%
        {Balsamo_Bajardi_De}
\bibfield{author}{\bibinfo{person}{Duilio Balsamo}, \bibinfo{person}{Paolo Bajardi}, \bibinfo{person}{Gianmarco De~Francisci~Morales}, \bibinfo{person}{Corrado Monti}, {and} \bibinfo{person}{Rossano Schifanella}.} \bibinfo{year}{2023}\natexlab{}.
\newblock \showarticletitle{The Pursuit of Peer Support for Opioid Use Recovery on Reddit}.
\newblock \bibinfo{journal}{\emph{Proceedings of the International AAAI Conference on Web and Social Media}} \bibinfo{volume}{17}, \bibinfo{number}{1} (\bibinfo{date}{Jun.} \bibinfo{year}{2023}), \bibinfo{pages}{12--23}.
\newblock
\urldef\tempurl%
\url{https://doi.org/10.1609/icwsm.v17i1.22122}
\showDOI{\tempurl}


\bibitem[Braun and Clarke(2006)]%
        {braun2006using}
\bibfield{author}{\bibinfo{person}{Virginia Braun} {and} \bibinfo{person}{Victoria Clarke}.} \bibinfo{year}{2006}\natexlab{}.
\newblock \showarticletitle{Using thematic analysis in psychology}.
\newblock \bibinfo{journal}{\emph{Qualitative research in psychology}} \bibinfo{volume}{3}, \bibinfo{number}{2} (\bibinfo{year}{2006}), \bibinfo{pages}{77--101}.
\newblock


\bibitem[Brown(1995)]%
        {brown1995naming}
\bibfield{author}{\bibinfo{person}{Phil Brown}.} \bibinfo{year}{1995}\natexlab{}.
\newblock \showarticletitle{Naming and framing: The social construction of diagnosis and illness}.
\newblock \bibinfo{journal}{\emph{Journal of health and social behavior}} (\bibinfo{year}{1995}), \bibinfo{pages}{34--52}.
\newblock


\bibitem[Chancellor et~al\mbox{.}(2019)]%
        {alternative_chi}
\bibfield{author}{\bibinfo{person}{Stevie Chancellor}, \bibinfo{person}{George Nitzburg}, \bibinfo{person}{Andrea Hu}, \bibinfo{person}{Francisco Zampieri}, {and} \bibinfo{person}{Munmun De~Choudhury}.} \bibinfo{year}{2019}\natexlab{}.
\newblock \showarticletitle{Discovering Alternative Treatments for Opioid Use Recovery Using Social Media}. In \bibinfo{booktitle}{\emph{Proceedings of the 2019 CHI Conference on Human Factors in Computing Systems}} (Glasgow, Scotland Uk) \emph{(\bibinfo{series}{CHI '19})}. \bibinfo{publisher}{Association for Computing Machinery}, \bibinfo{address}{New York, NY, USA}, \bibinfo{pages}{1–15}.
\newblock
\showISBNx{9781450359702}
\urldef\tempurl%
\url{https://doi.org/10.1145/3290605.3300354}
\showDOI{\tempurl}


\bibitem[Chandra and Pal(2019)]%
        {10.1145/3290605.3300563}
\bibfield{author}{\bibinfo{person}{Priyank Chandra} {and} \bibinfo{person}{Joyojeet Pal}.} \bibinfo{year}{2019}\natexlab{}.
\newblock \showarticletitle{Rumors and Collective Sensemaking: Managing Ambiguity in an Informal Marketplace}. In \bibinfo{booktitle}{\emph{Proceedings of the 2019 CHI Conference on Human Factors in Computing Systems}} (Glasgow, Scotland Uk) \emph{(\bibinfo{series}{CHI '19})}. \bibinfo{publisher}{Association for Computing Machinery}, \bibinfo{address}{New York, NY, USA}, \bibinfo{pages}{1–12}.
\newblock
\showISBNx{9781450359702}
\urldef\tempurl%
\url{https://doi.org/10.1145/3290605.3300563}
\showDOI{\tempurl}


\bibitem[Charmaz(1990)]%
        {charmaz1990discovering}
\bibfield{author}{\bibinfo{person}{Kathy Charmaz}.} \bibinfo{year}{1990}\natexlab{}.
\newblock \showarticletitle{‘Discovering’chronic illness: using grounded theory}.
\newblock \bibinfo{journal}{\emph{Social science \& medicine}} \bibinfo{volume}{30}, \bibinfo{number}{11} (\bibinfo{year}{1990}), \bibinfo{pages}{1161--1172}.
\newblock


\bibitem[Chen et~al\mbox{.}(2022)]%
        {CHEN2022100061}
\bibfield{author}{\bibinfo{person}{Annie~T. Chen}, \bibinfo{person}{Shana Johnny}, {and} \bibinfo{person}{Mike Conway}.} \bibinfo{year}{2022}\natexlab{}.
\newblock \showarticletitle{Examining stigma relating to substance use and contextual factors in social media discussions}.
\newblock \bibinfo{journal}{\emph{Drug and Alcohol Dependence Reports}}  \bibinfo{volume}{3} (\bibinfo{year}{2022}), \bibinfo{pages}{100061}.
\newblock
\showISSN{2772-7246}
\urldef\tempurl%
\url{https://doi.org/10.1016/j.dadr.2022.100061}
\showDOI{\tempurl}


\bibitem[Chen and Shu(2023)]%
        {chen2023combatingmisinformationagellms}
\bibfield{author}{\bibinfo{person}{Canyu Chen} {and} \bibinfo{person}{Kai Shu}.} \bibinfo{year}{2023}\natexlab{}.
\newblock \bibinfo{title}{Combating Misinformation in the Age of LLMs: Opportunities and Challenges}.
\newblock
\newblock
\showeprint[arxiv]{2311.05656}~[cs.CY]
\urldef\tempurl%
\url{https://arxiv.org/abs/2311.05656}
\showURL{%
\tempurl}


\bibitem[Choi et~al\mbox{.}(2023)]%
        {choi-etal-2023-llms}
\bibfield{author}{\bibinfo{person}{Minje Choi}, \bibinfo{person}{Jiaxin Pei}, \bibinfo{person}{Sagar Kumar}, \bibinfo{person}{Chang Shu}, {and} \bibinfo{person}{David Jurgens}.} \bibinfo{year}{2023}\natexlab{}.
\newblock \showarticletitle{Do {LLM}s Understand Social Knowledge? Evaluating the Sociability of Large Language Models with {S}oc{KET} Benchmark}. In \bibinfo{booktitle}{\emph{Proceedings of the 2023 Conference on Empirical Methods in Natural Language Processing}}, \bibfield{editor}{\bibinfo{person}{Houda Bouamor}, \bibinfo{person}{Juan Pino}, {and} \bibinfo{person}{Kalika Bali}} (Eds.). \bibinfo{publisher}{Association for Computational Linguistics}, \bibinfo{address}{Singapore}, \bibinfo{pages}{11370--11403}.
\newblock
\urldef\tempurl%
\url{https://doi.org/10.18653/v1/2023.emnlp-main.699}
\showDOI{\tempurl}


\bibitem[Colloca and Miller(2011)]%
        {colloca2011nocebo}
\bibfield{author}{\bibinfo{person}{Luana Colloca} {and} \bibinfo{person}{Franklin~G Miller}.} \bibinfo{year}{2011}\natexlab{}.
\newblock \showarticletitle{The nocebo effect and its relevance for clinical practice}.
\newblock \bibinfo{journal}{\emph{Psychosomatic medicine}} \bibinfo{volume}{73}, \bibinfo{number}{7} (\bibinfo{year}{2011}), \bibinfo{pages}{598--603}.
\newblock


\bibitem[Conrad and Barker(2010)]%
        {conrad2010social}
\bibfield{author}{\bibinfo{person}{Peter Conrad} {and} \bibinfo{person}{Kristin~K Barker}.} \bibinfo{year}{2010}\natexlab{}.
\newblock \showarticletitle{The social construction of illness: Key insights and policy implications}.
\newblock \bibinfo{journal}{\emph{Journal of health and social behavior}} \bibinfo{volume}{51}, \bibinfo{number}{1\_suppl} (\bibinfo{year}{2010}), \bibinfo{pages}{S67--S79}.
\newblock


\bibitem[Dickson-Gomez et~al\mbox{.}(2022)]%
        {dickson2022you}
\bibfield{author}{\bibinfo{person}{Julia Dickson-Gomez}, \bibinfo{person}{Antoinette Spector}, \bibinfo{person}{Margaret Weeks}, \bibinfo{person}{Carol Galletly}, \bibinfo{person}{Madelyn McDonald}, {and} \bibinfo{person}{Helena~Danielle Green~Montaque}.} \bibinfo{year}{2022}\natexlab{}.
\newblock \showarticletitle{“You’re not supposed to be on it forever”: Medications to treat opioid use disorder (MOUD) related stigma among drug treatment providers and people who use opioids}.
\newblock \bibinfo{journal}{\emph{Substance Abuse: Research and Treatment}}  \bibinfo{volume}{16} (\bibinfo{year}{2022}), \bibinfo{pages}{11782218221103859}.
\newblock


\bibitem[Dydyk et~al\mbox{.}(2023)]%
        {dydyk2023opioid}
\bibfield{author}{\bibinfo{person}{Alexander~M Dydyk}, \bibinfo{person}{Nitesh~K Jain}, {and} \bibinfo{person}{Mohit Gupta}.} \bibinfo{year}{2023}\natexlab{}.
\newblock \showarticletitle{Opioid use disorder}.
\newblock In \bibinfo{booktitle}{\emph{StatPearls [Internet]}}. \bibinfo{publisher}{StatPearls Publishing}.
\newblock


\bibitem[Farhana et~al\mbox{.}(2023)]%
        {farhana2023social}
\bibfield{author}{\bibinfo{person}{Nusrat Farhana}, \bibinfo{person}{Allie Peckham}, \bibinfo{person}{Husayn Marani}, \bibinfo{person}{Monika Roerig}, {and} \bibinfo{person}{Greg Marchildon}.} \bibinfo{year}{2023}\natexlab{}.
\newblock \showarticletitle{The Social Construction of Dementia: Implications for Healthcare Experiences of Caregivers and People Living with Dementia}.
\newblock \bibinfo{journal}{\emph{Journal of Patient Experience}}  \bibinfo{volume}{10} (\bibinfo{year}{2023}), \bibinfo{pages}{23743735231211066}.
\newblock


\bibitem[FDA(2024)]%
        {FDA}
FDA \bibinfo{year}{2024}\natexlab{}.
\newblock \bibinfo{booktitle}{\emph{U.S. Food \& Drug Administration}}.
\newblock
\urldef\tempurl%
\url{https://www.fda.gov/}
\showURL{%
\tempurl}
\newblock
\shownote{Last Accessed: 2024-08-10}.


\bibitem[Florence et~al\mbox{.}(2021)]%
        {FLORENCE2021108350}
\bibfield{author}{\bibinfo{person}{Curtis Florence}, \bibinfo{person}{Feijun Luo}, {and} \bibinfo{person}{Ketra Rice}.} \bibinfo{year}{2021}\natexlab{}.
\newblock \showarticletitle{The economic burden of opioid use disorder and fatal opioid overdose in the United States, 2017}.
\newblock \bibinfo{journal}{\emph{Drug and Alcohol Dependence}}  \bibinfo{volume}{218} (\bibinfo{year}{2021}), \bibinfo{pages}{108350}.
\newblock
\showISSN{0376-8716}
\urldef\tempurl%
\url{https://doi.org/10.1016/j.drugalcdep.2020.108350}
\showDOI{\tempurl}


\bibitem[Househ(2013)]%
        {househ2013use}
\bibfield{author}{\bibinfo{person}{Mowafa Househ}.} \bibinfo{year}{2013}\natexlab{}.
\newblock \showarticletitle{The use of social media in healthcare: organizational, clinical, and patient perspectives}.
\newblock \bibinfo{journal}{\emph{Enabling health and healthcare through ICT}} (\bibinfo{year}{2013}), \bibinfo{pages}{244--248}.
\newblock


\bibitem[Kanchan and Gaidhane(2023)]%
        {kanchan2023social}
\bibfield{author}{\bibinfo{person}{Sushim Kanchan} {and} \bibinfo{person}{Abhay Gaidhane}.} \bibinfo{year}{2023}\natexlab{}.
\newblock \showarticletitle{Social media role and its impact on public health: A narrative review}.
\newblock \bibinfo{journal}{\emph{Cureus}} \bibinfo{volume}{15}, \bibinfo{number}{1} (\bibinfo{year}{2023}).
\newblock


\bibitem[Kim et~al\mbox{.}(2024)]%
        {10.1145/3613904.3642937}
\bibfield{author}{\bibinfo{person}{Taewan Kim}, \bibinfo{person}{Seolyeong Bae}, \bibinfo{person}{Hyun~Ah Kim}, \bibinfo{person}{Su-Woo Lee}, \bibinfo{person}{Hwajung Hong}, \bibinfo{person}{Chanmo Yang}, {and} \bibinfo{person}{Young-Ho Kim}.} \bibinfo{year}{2024}\natexlab{}.
\newblock \showarticletitle{MindfulDiary: Harnessing Large Language Model to Support Psychiatric Patients' Journaling}. In \bibinfo{booktitle}{\emph{Proceedings of the CHI Conference on Human Factors in Computing Systems}} (Honolulu, HI, USA) \emph{(\bibinfo{series}{CHI '24})}. \bibinfo{publisher}{Association for Computing Machinery}, \bibinfo{address}{New York, NY, USA}, Article \bibinfo{articleno}{701}, \bibinfo{numpages}{20}~pages.
\newblock
\showISBNx{9798400703300}
\urldef\tempurl%
\url{https://doi.org/10.1145/3613904.3642937}
\showDOI{\tempurl}


\bibitem[Lavertu et~al\mbox{.}(2021)]%
        {Lavertu2021.04.01.21254815}
\bibfield{author}{\bibinfo{person}{Adam Lavertu}, \bibinfo{person}{Tymor Hamamsy}, {and} \bibinfo{person}{Russ~B Altman}.} \bibinfo{year}{2021}\natexlab{}.
\newblock \showarticletitle{Monitoring the opioid epidemic via social media discussions}.
\newblock \bibinfo{journal}{\emph{medRxiv}} (\bibinfo{year}{2021}).
\newblock
\urldef\tempurl%
\url{https://doi.org/10.1101/2021.04.01.21254815}
\showDOI{\tempurl}
\showeprint{https://www.medrxiv.org/content/early/2021/04/07/2021.04.01.21254815.full.pdf}


\bibitem[Lee and Wu(2014)]%
        {lee2014effects}
\bibfield{author}{\bibinfo{person}{Yi-Chih Lee} {and} \bibinfo{person}{Wei-Li Wu}.} \bibinfo{year}{2014}\natexlab{}.
\newblock \showarticletitle{The effects of situated learning and health knowledge involvement on health communications}.
\newblock \bibinfo{journal}{\emph{Reproductive health}}  \bibinfo{volume}{11} (\bibinfo{year}{2014}), \bibinfo{pages}{1--7}.
\newblock


\bibitem[Li et~al\mbox{.}(2014)]%
        {li-etal-2014-major}
\bibfield{author}{\bibinfo{person}{Jiwei Li}, \bibinfo{person}{Alan Ritter}, \bibinfo{person}{Claire Cardie}, {and} \bibinfo{person}{Eduard Hovy}.} \bibinfo{year}{2014}\natexlab{}.
\newblock \showarticletitle{Major Life Event Extraction from {T}witter based on Congratulations/Condolences Speech Acts}. In \bibinfo{booktitle}{\emph{Proceedings of the 2014 Conference on Empirical Methods in Natural Language Processing ({EMNLP})}}. \bibinfo{address}{Doha, Qatar}, \bibinfo{pages}{1997--2007}.
\newblock
\urldef\tempurl%
\url{https://doi.org/10.3115/v1/D14-1214}
\showDOI{\tempurl}


\bibitem[Lupton(2012)]%
        {lupton2012medicine}
\bibfield{author}{\bibinfo{person}{Deborah Lupton}.} \bibinfo{year}{2012}\natexlab{}.
\newblock \showarticletitle{Medicine as culture: illness, disease and the body}.
\newblock  (\bibinfo{year}{2012}).
\newblock


\bibitem[Maines(2000)]%
        {maines2000social}
\bibfield{author}{\bibinfo{person}{David~R Maines}.} \bibinfo{year}{2000}\natexlab{}.
\newblock \showarticletitle{The social construction of meaning}.
\newblock \bibinfo{journal}{\emph{Contemporary Sociology}} \bibinfo{volume}{29}, \bibinfo{number}{4} (\bibinfo{year}{2000}), \bibinfo{pages}{577--584}.
\newblock


\bibitem[Mamykina et~al\mbox{.}(2015)]%
        {10.1145/2702123.2702566}
\bibfield{author}{\bibinfo{person}{Lena Mamykina}, \bibinfo{person}{Drashko Nakikj}, {and} \bibinfo{person}{Noemie Elhadad}.} \bibinfo{year}{2015}\natexlab{}.
\newblock \showarticletitle{Collective Sensemaking in Online Health Forums}. In \bibinfo{booktitle}{\emph{Proceedings of the 33rd Annual ACM Conference on Human Factors in Computing Systems}} (Seoul, Republic of Korea) \emph{(\bibinfo{series}{CHI '15})}. \bibinfo{publisher}{Association for Computing Machinery}, \bibinfo{address}{New York, NY, USA}, \bibinfo{pages}{3217–3226}.
\newblock
\showISBNx{9781450331456}
\urldef\tempurl%
\url{https://doi.org/10.1145/2702123.2702566}
\showDOI{\tempurl}


\bibitem[Martin and Peterson(2009)]%
        {martin2009social}
\bibfield{author}{\bibinfo{person}{Carmel~M Martin} {and} \bibinfo{person}{Chris Peterson}.} \bibinfo{year}{2009}\natexlab{}.
\newblock \showarticletitle{The social construction of chronicity--a key to understanding chronic care transformations}.
\newblock \bibinfo{journal}{\emph{Journal of Evaluation in Clinical Practice}} \bibinfo{volume}{15}, \bibinfo{number}{3} (\bibinfo{year}{2009}), \bibinfo{pages}{578--585}.
\newblock


\bibitem[Metzger et~al\mbox{.}(2010)]%
        {metzger2010social}
\bibfield{author}{\bibinfo{person}{Miriam~J Metzger}, \bibinfo{person}{Andrew~J Flanagin}, {and} \bibinfo{person}{Ryan~B Medders}.} \bibinfo{year}{2010}\natexlab{}.
\newblock \showarticletitle{Social and heuristic approaches to credibility evaluation online}.
\newblock \bibinfo{journal}{\emph{Journal of communication}} \bibinfo{volume}{60}, \bibinfo{number}{3} (\bibinfo{year}{2010}), \bibinfo{pages}{413--439}.
\newblock


\bibitem[Min et~al\mbox{.}(2022)]%
        {min-etal-2022-rethinking}
\bibfield{author}{\bibinfo{person}{Sewon Min}, \bibinfo{person}{Xinxi Lyu}, \bibinfo{person}{Ari Holtzman}, \bibinfo{person}{Mikel Artetxe}, \bibinfo{person}{Mike Lewis}, \bibinfo{person}{Hannaneh Hajishirzi}, {and} \bibinfo{person}{Luke Zettlemoyer}.} \bibinfo{year}{2022}\natexlab{}.
\newblock \showarticletitle{Rethinking the Role of Demonstrations: What Makes In-Context Learning Work?}. In \bibinfo{booktitle}{\emph{Proceedings of the 2022 Conference on Empirical Methods in Natural Language Processing}}, \bibfield{editor}{\bibinfo{person}{Yoav Goldberg}, \bibinfo{person}{Zornitsa Kozareva}, {and} \bibinfo{person}{Yue Zhang}} (Eds.). \bibinfo{publisher}{Association for Computational Linguistics}, \bibinfo{address}{Abu Dhabi, United Arab Emirates}, \bibinfo{pages}{11048--11064}.
\newblock
\urldef\tempurl%
\url{https://doi.org/10.18653/v1/2022.emnlp-main.759}
\showDOI{\tempurl}


\bibitem[Mistry et~al\mbox{.}(2015)]%
        {mistry2015technology}
\bibfield{author}{\bibinfo{person}{Niraj Mistry}, \bibinfo{person}{Arun Keepanasseril}, \bibinfo{person}{Nancy~L Wilczynski}, \bibinfo{person}{Robby Nieuwlaat}, \bibinfo{person}{Manthan Ravall}, \bibinfo{person}{R~Brian Haynes}, {and} \bibinfo{person}{Patient Adherence~Review Team}.} \bibinfo{year}{2015}\natexlab{}.
\newblock \showarticletitle{Technology-mediated interventions for enhancing medication adherence}.
\newblock \bibinfo{journal}{\emph{Journal of the American Medical Informatics Association}} \bibinfo{volume}{22}, \bibinfo{number}{e1} (\bibinfo{year}{2015}), \bibinfo{pages}{e177--e193}.
\newblock


\bibitem[{OpenAI}(2020)]%
        {openai_reddit_partnership_2024}
\bibfield{author}{\bibinfo{person}{{OpenAI}}.} \bibinfo{year}{2020}\natexlab{}.
\newblock \bibinfo{title}{OpenAI and Reddit Partnership}.
\newblock \bibinfo{howpublished}{\url{https://openai.com/index/openai-and-reddit-partnership/}}.
\newblock
\newblock
\shownote{Accessed: 2024-08-10}.


\bibitem[OpenAI(2024)]%
        {openai2024gpt4technicalreport}
\bibfield{author}{\bibinfo{person}{OpenAI}.} \bibinfo{year}{2024}\natexlab{}.
\newblock \bibinfo{title}{GPT-4 Technical Report}.
\newblock
\newblock
\showeprint[arxiv]{2303.08774}~[cs.CL]
\urldef\tempurl%
\url{https://arxiv.org/abs/2303.08774}
\showURL{%
\tempurl}


\bibitem[Papoutsaki et~al\mbox{.}(2021)]%
        {10.1145/3479564}
\bibfield{author}{\bibinfo{person}{Alexandra Papoutsaki}, \bibinfo{person}{Samuel So}, \bibinfo{person}{Georgia Kenderova}, \bibinfo{person}{Bryan Shapiro}, {and} \bibinfo{person}{Daniel~A. Epstein}.} \bibinfo{year}{2021}\natexlab{}.
\newblock \showarticletitle{Understanding Delivery of Collectively Built Protocols in an Online Health Community for Discontinuation of Psychiatric Drugs}.
\newblock \bibinfo{journal}{\emph{Proc. ACM Hum.-Comput. Interact.}} \bibinfo{volume}{5}, \bibinfo{number}{CSCW2}, Article \bibinfo{articleno}{420} (\bibinfo{date}{oct} \bibinfo{year}{2021}), \bibinfo{numpages}{29}~pages.
\newblock
\urldef\tempurl%
\url{https://doi.org/10.1145/3479564}
\showDOI{\tempurl}


\bibitem[PCSS(2024)]%
        {PCSS}
PCSS \bibinfo{year}{2024}\natexlab{}.
\newblock \bibinfo{booktitle}{\emph{Providers Clinical Support System (PCSS)}}.
\newblock
\urldef\tempurl%
\url{https://pcssnow.org/}
\showURL{%
\tempurl}
\newblock
\shownote{Last Accessed: 2024-08-10}.


\bibitem[Pinch and Bijker(1984)]%
        {pinch1984social}
\bibfield{author}{\bibinfo{person}{Trevor~J Pinch} {and} \bibinfo{person}{Wiebe~E Bijker}.} \bibinfo{year}{1984}\natexlab{}.
\newblock \showarticletitle{The social construction of facts and artefacts: Or how the sociology of science and the sociology of technology might benefit each other}.
\newblock \bibinfo{journal}{\emph{Social studies of science}} \bibinfo{volume}{14}, \bibinfo{number}{3} (\bibinfo{year}{1984}), \bibinfo{pages}{399--441}.
\newblock


\bibitem[Pratiwi et~al\mbox{.}(2023)]%
        {pratiwi2023systematic}
\bibfield{author}{\bibinfo{person}{Hening Pratiwi}, \bibinfo{person}{Susi~Ari Kristina}, \bibinfo{person}{Anna~Wahyuni Widayanti}, \bibinfo{person}{Yayi~Suryo Prabandari}, {and} \bibinfo{person}{Ikhwan~Yuda Kusuma}.} \bibinfo{year}{2023}\natexlab{}.
\newblock \showarticletitle{A systematic review of compensation and technology-mediated strategies to maintain older adults’ medication adherence}.
\newblock \bibinfo{journal}{\emph{International Journal of Environmental Research and Public Health}} \bibinfo{volume}{20}, \bibinfo{number}{1} (\bibinfo{year}{2023}), \bibinfo{pages}{803}.
\newblock


\bibitem[SAMHSA(2024)]%
        {SAMSHA}
SAMHSA \bibinfo{year}{2024}\natexlab{}.
\newblock \bibinfo{booktitle}{\emph{Substance Abuse and Mental Health Services Administration (SAMHSA)}}.
\newblock
\urldef\tempurl%
\url{https://www.samhsa.gov/}
\showURL{%
\tempurl}
\newblock
\shownote{Last Accessed: 2024-08-10}.


\bibitem[Sharif et~al\mbox{.}(2024)]%
        {sharif2024characterizing}
\bibfield{author}{\bibinfo{person}{Omar Sharif}, \bibinfo{person}{Madhusudan Basak}, \bibinfo{person}{Tanzia Parvin}, \bibinfo{person}{Ava Scharfstein}, \bibinfo{person}{Alphonso Bradham}, \bibinfo{person}{Jacob~T Borodovsky}, \bibinfo{person}{Sarah~E Lord}, {and} \bibinfo{person}{Sarah~M Preum}.} \bibinfo{year}{2024}\natexlab{}.
\newblock \showarticletitle{Characterizing Information Seeking Events in Health-Related Social Discourse}. In \bibinfo{booktitle}{\emph{Proceedings of the AAAI Conference on Artificial Intelligence}}, Vol.~\bibinfo{volume}{38}. \bibinfo{pages}{22350--22358}.
\newblock


\bibitem[Spiro and Starbird(2023)]%
        {spiro2023rumors}
\bibfield{author}{\bibinfo{person}{Emma Spiro} {and} \bibinfo{person}{Kate Starbird}.} \bibinfo{year}{2023}\natexlab{}.
\newblock \showarticletitle{Rumors have rules}.
\newblock \bibinfo{journal}{\emph{Issues in Science and Technology}} \bibinfo{volume}{29}, \bibinfo{number}{3} (\bibinfo{year}{2023}).
\newblock


\bibitem[Suboxone.com(2024)]%
        {Suboxone}
Suboxone.com \bibinfo{year}{2024}\natexlab{}.
\newblock \bibinfo{booktitle}{\emph{Suboxone.com}}.
\newblock
\urldef\tempurl%
\url{https://www.suboxone.com/}
\showURL{%
\tempurl}
\newblock
\shownote{Last Accessed: 2024-08-10}.


\bibitem[Timmermans and Tavory(2012)]%
        {timmermans2012theory}
\bibfield{author}{\bibinfo{person}{Stefan Timmermans} {and} \bibinfo{person}{Iddo Tavory}.} \bibinfo{year}{2012}\natexlab{}.
\newblock \showarticletitle{Theory construction in qualitative research: From grounded theory to abductive analysis}.
\newblock \bibinfo{journal}{\emph{Sociological theory}} \bibinfo{volume}{30}, \bibinfo{number}{3} (\bibinfo{year}{2012}), \bibinfo{pages}{167--186}.
\newblock


\bibitem[Vaishya et~al\mbox{.}(2023)]%
        {vaishya2023chatgpt}
\bibfield{author}{\bibinfo{person}{Raju Vaishya}, \bibinfo{person}{Anoop Misra}, {and} \bibinfo{person}{Abhishek Vaish}.} \bibinfo{year}{2023}\natexlab{}.
\newblock \showarticletitle{ChatGPT: Is this version good for healthcare and research?}
\newblock \bibinfo{journal}{\emph{Diabetes \& Metabolic Syndrome: Clinical Research \& Reviews}} \bibinfo{volume}{17}, \bibinfo{number}{4} (\bibinfo{year}{2023}), \bibinfo{pages}{102744}.
\newblock


\bibitem[Ziems et~al\mbox{.}(2024)]%
        {ziems-etal-2024-large}
\bibfield{author}{\bibinfo{person}{Caleb Ziems}, \bibinfo{person}{William Held}, \bibinfo{person}{Omar Shaikh}, \bibinfo{person}{Jiaao Chen}, \bibinfo{person}{Zhehao Zhang}, {and} \bibinfo{person}{Diyi Yang}.} \bibinfo{year}{2024}\natexlab{}.
\newblock \showarticletitle{Can Large Language Models Transform Computational Social Science?}
\newblock \bibinfo{journal}{\emph{Computational Linguistics}} \bibinfo{volume}{50}, \bibinfo{number}{1} (\bibinfo{date}{March} \bibinfo{year}{2024}), \bibinfo{pages}{237--291}.
\newblock
\urldef\tempurl%
\url{https://doi.org/10.1162/coli_a_00502}
\showDOI{\tempurl}


\end{thebibliography}
